\documentclass[lettersize,journal]{IEEEtran}
\usepackage{cite}
\usepackage{amsmath,amssymb,amsfonts}
\usepackage{algorithm}
\usepackage{algpseudocode}
\usepackage{graphicx}
\usepackage{textcomp}
\usepackage{xcolor}
\usepackage{enumitem} 
\graphicspath{ {images/} }
\usepackage{makecell} 
\usepackage{bm}
\usepackage{booktabs}
\usepackage{multirow}
\begin{document}
	
	\title{Network-Aware Progressive Replacement Sequence Planning for Laser-Enhanced BeiDou Navigation Constellations}
	
		\author{
		\IEEEauthorblockN{
			Huan Yan\IEEEauthorrefmark{1},
			Juan A. FRAIRE\IEEEauthorrefmark{2}\IEEEauthorrefmark{3},
			Ziqi Yang\IEEEauthorrefmark{1},
			Kanglian Zhao\IEEEauthorrefmark{1},
			Wenfeng Li\IEEEauthorrefmark{1},
			Jinjun Zheng\IEEEauthorrefmark{4},
			Ziyu Wang\IEEEauthorrefmark{4},
			Changgang Zheng\IEEEauthorrefmark{4},
		}
		
		\IEEEauthorblockA{
			\IEEEauthorrefmark{1} Nanjing University, Nanjing, China \\
			\IEEEauthorrefmark{2} Inria, INSA Lyon, CITI, UR3720, 69621 Villeurbanne, France\\
			\IEEEauthorrefmark{3} CONICET -- Universidad Nacional de Córdoba, Córdoba, Argentina \\
			\IEEEauthorrefmark{4} China Academy of Space Technology, Beijing, China \\
		}
		\thanks{(Corresponding authors: Kanglian Zhao; Jinjun Zheng.)}
	}

	\maketitle
	
\begin{abstract}
	The early satellites of the third-generation BeiDou Navigation Satellite System
	(BDS-3) are approaching the end of their design lifetime, making progressive
	constellation replacement an inevitable engineering task. Meanwhile, laser
	inter-satellite links (LISLs) offer high-precision time transfer, ranging, and
	high-rate data forwarding, providing a promising upgrade path for future
	BeiDou satellites. This paper investigates the replacement-sequence planning
	problem for progressively replacing legacy microwave satellites with
	laser-enabled satellites from a networking perspective.
	Since practical replacement decisions remain constrained by
	legacy satellite lifetime, on-orbit condition, laser satellite manufacturing readiness, and launch scheduling, networking
	information is used as an auxiliary criterion in the actual engineering replacement process.
	An integer linear programming model is
	developed to evaluate the network value of intermediate laser--microwave hybrid
	states. Based on this common state evaluator, two
	complementary planning methods are developed. A priority-aware exact method
	supports round-wise network-assisted launch decisions, while a
	visibility-guided adaptive large-neighborhood search is used for full-sequence
	planning and provides a full-horizon network-oriented near-optimal reference.
	Simulations using the actual BDS-3 constellation show that stronger network
	participation leads to earlier and larger ground-connected laser subnetworks,
	fewer microwave-dependent satellites, better coordination between laser-backbone
	growth and high-rate ground access, and improved residual microwave-layer
	performance. These results provide a quantitative basis for incorporating
	network performance as an auxiliary criterion in future BeiDou replacement
	planning.
\end{abstract}
	

	
	
\section{Introduction}
\label{sec:introduction}
In 2017, China launched the first satellite of the third-generation BeiDou Navigation Satellite System (BDS-3)~\cite{1}. Since the design lifetime of a BDS-3 satellite is approximately ten years~\cite{2}, the early satellites in the current constellation are now approaching the end of their service life and must be replaced by newly launched satellites.

Meanwhile, a growing body of research has investigated the application of laser inter-satellite links (LISLs) to future navigation constellations. Compared with the Ka-band phased-array microwave ISLs currently adopted in BeiDou~\cite{2,9}, LISLs offer several prominent advantages. First, laser links can achieve picosecond-level time transfer accuracy~\cite{3,8}, thereby significantly improving the timing performance of the system. Second, the continuous sub-millimeter ranging capability of laser links can provide a large amount of high-precision inter-satellite measurement data~\cite{4,6}, which in turn improves orbit determination accuracy and enables better GNSS positioning services for users~\cite{5}. Third, the Gbps-level transmission rate of laser links enables high-speed inter-satellite networking and large-volume data forwarding and exchange, thus substantially enhancing the communication and network management capability of the system~\cite{6,7}. 

Motivated by these facts, the next major evolution of the BDS-3 is to progressively replace the legacy microwave satellites reaching the end of their lifetime with newly launched satellites carrying laser payloads, hereafter referred to as \emph{laser satellites}~\cite{10,11}. It should be noted, however, that laser payloads do not eliminate the need for microwave payloads. In order to support diversified inter-satellite measurement tasks, navigation satellites must establish links and perform ranging with different objects within a short time. Because laser links require sophisticated pointing, acquisition, and tracking (PAT), their link switching process is slow and incurs a high reconfiguration cost~\cite{10}. Therefore, even the newly launched laser satellites still need to be equipped with fast-switching microwave phased-array antennas. In this sense, the laser payload is an incremental enhancement rather than a complete replacement of the microwave payload.

Most existing studies focus on topology construction for LISLs in navigation constellations and typically optimize global metrics such as Position Dilution of Precision (PDOP)~\cite{11}, end-to-end network delay~\cite{12}, and network-wide connectivity~\cite{10}. However, these studies usually assume that the laser-enabled satellites have already been fully or partially deployed as a given set, while overlooking the gradual replacement process itself. In the early replacement stage, the number of deployed laser satellites is still limited, the resulting laser topology may be disconnected, and many launched laser satellites may remain outside the connected laser part of the constellation. Under such conditions, conventional metrics such as PDOP and network-wide connectivity are no longer suitable for guiding launch planning.
Moreover, replacing an existing navigation constellation is fundamentally different from deploying a new constellation from scratch. During the initial deployment of BDS-3, the primary objective was to ensure basic Radio Navigation Satellite Service (RNSS) capability, including global coverage and geometric indicators such as PDOP~\cite{9}. 
In contrast, in the replacement problem studied in this paper, the original microwave constellation is already operational. The system therefore evolves from a functionally complete but capability-limited architecture to a stronger laser--microwave hybrid architecture, rather than from no service to full service. Consequently, the key planning problem is no longer how to establish the basic functionality of the constellation, but how to optimize the intermediate states generated during a multi-year replacement process.
The historical deployment of BDS-3 itself lasted from 2017 to 2020~\cite{1,2}. Likewise, the progressive replacement process considered here also spans multiple rounds separated by production scheduling, launch-site coordination, and mission preparation. The replacement order therefore directly affects the networking quality of the intermediate laser--microwave hybrid constellation over each inter-launch interval.

\begin{figure}[t]
	\centering
	\includegraphics[width=0.75\linewidth]{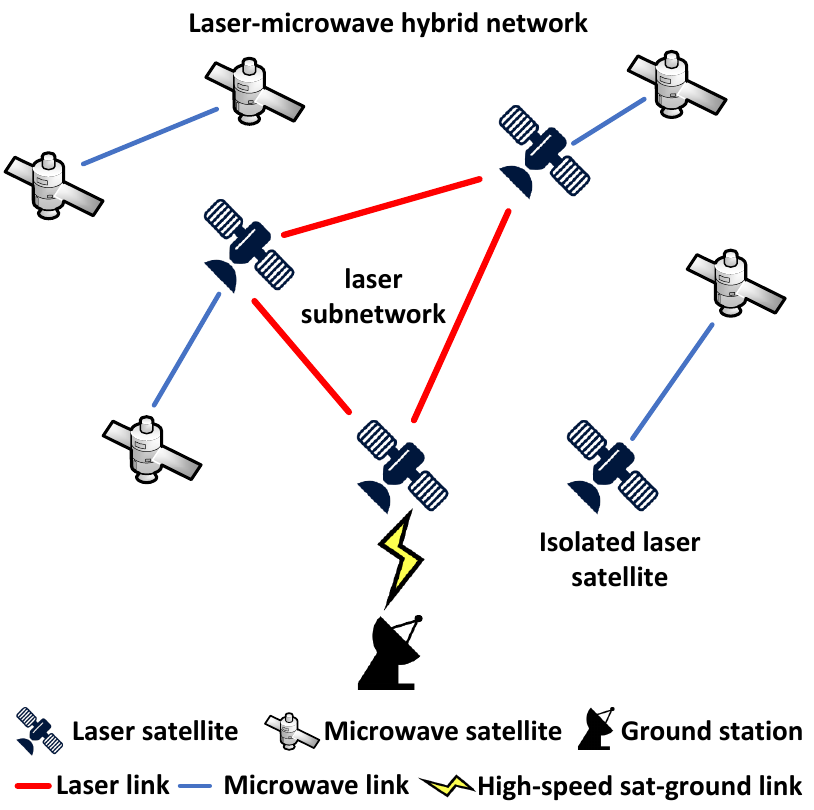}
	\caption{Intermediate laser--microwave hybrid constellation during progressive replacement. A ground-accessible local laser-subnet coexists with residual microwave satellites, while some launched laser satellites may remain isolated.}
	\label{fig:hybrid_scenario}
\end{figure}

Fig.~\ref{fig:hybrid_scenario} illustrates the intermediate hybrid constellation state during progressive replacement. In this state, a ground-accessible local laser subnetwork coexists with residual microwave satellites, while some launched laser satellites may remain outside the connected laser subnetwork because of visibility limitations and terminal-resource constraints. Such laser--microwave coexistence defines the networking scenario to be optimized during replacement.

Based on the above observations, this paper investigates the launch-sequence
planning problem for progressively replacing legacy microwave satellites with
laser-enabled satellites in the BeiDou navigation constellation from a
networking perspective. In practical constellation replacement, launch
decisions are jointly influenced by legacy satellite lifetime, on-orbit
condition, laser satellite manufacturing readiness, launch-task scheduling, and other
engineering considerations. Network performance constitutes an auxiliary
decision dimension in this process. This paper therefore focuses on
characterizing the influence of networking information on replacement-sequence
planning and quantifying the network benefits obtained when such information
is progressively incorporated into launch decisions. The resulting framework
provides a basis for including network performance as an auxiliary criterion
in future engineering replacement planning.

To characterize the network value of different intermediate replacement
states, we introduce a common \emph{networking potential} measure and formulate
its evaluation as a single-commodity-flow integer linear programming
(SCF-ILP) problem. Based on this common state evaluator, two complementary
planning modes are developed. Priority-aware exact round-wise planning supports
network-assisted decisions at individual launch opportunities as engineering
conditions evolve, while visibility-guided adaptive large-neighborhood search
(ALNS) explores complete replacement sequences and provides a near-optimal
full-horizon network-oriented reference. The common state evaluator, together
with incremental reoptimization and deployed-state reuse, enables efficient evaluation across both planning modes.
The main contributions of this paper are summarized as follows:
\begin{itemize}[leftmargin=*]
	\item The progressive replacement of the BeiDou navigation constellation is
	formulated as a network-aware launch-sequence planning problem over evolving
	laser--microwave hybrid states. A common networking potential measure is
	introduced to characterize the network value of each intermediate deployed
	state, and a compact SCF-ILP model is developed to compute this potential
	under practical launch and network-resource conditions. 
	
	\item Two complementary planning methods are developed on the common
	state-evaluation model. Priority-aware exact round-wise planning supports
	network-assisted decisions at individual launch opportunities, while
	visibility-guided ALNS searches the complete replacement trajectory and
	provides a near-optimal full-horizon sequence reference.
	
	\item Simulations based on the actual BDS-3 constellation progressively
	increase the participation of network information in launch decisions. The
	results show that stronger network participation leads to earlier and larger
	ground-connected laser subnetworks, fewer residual microwave-dependent
	satellites, better coordination between laser-backbone growth and high-rate
	ground access, and improved ranging and waiting-delay performance in the
	residual microwave layer.
\end{itemize}

The remainder of this paper is organized as follows.
Section~\ref{sec:system_model} models the progressive replacement process,
hybrid-network operation, and networking potential.
Section~\ref{sec:potential_computation} develops the networking potential
computation and representative topology realization.
Section~\ref{sec:replacement_planning_methods} presents the round-wise and
full-sequence planning methods.
Section~\ref{sec:evaluation} evaluates computational efficiency and the
network-performance evolution under different levels of network participation.
Finally, Section~\ref{sec:conclusion} concludes this paper.

\section{Progressive Replacement Model and Networking Potential}
\label{sec:system_model}

Progressive replacement proceeds through a sequence of laser--microwave hybrid constellation states. 
This section models the replacement-state evolution, describes the operating environment of the hybrid network, defines the networking potential of an intermediate deployment state, and formulates the round-wise and full-sequence planning problems.

\subsection{Replacement-Candidate Satellites and Launch-State Evolution}
\label{subsec:replacement_state}

Let $\mathcal{S}$ denote the set of satellites involved in the progressive replacement process, and let $r=1,2,\ldots,R$ index the launch rounds. After round $r$, the positions occupied by laser-enabled replacement satellites form the deployed set $\mathcal{H}_r\subseteq\mathcal{S}$, with $\mathcal{H}_0$ denoting the initial deployed set. The positions in $\mathcal{S}\setminus\mathcal{H}_r$ continue to be occupied by legacy microwave satellites.

At each launch opportunity, engineering considerations such as legacy microwave satellite on-orbit condition, remaining lifetime, laser satellite manufacturing readiness, and launch-task scheduling determine the satellites that preferably enter replacement consideration. Let
\begin{equation}
	\mathcal{C}_r
	\subseteq
	\mathcal{S}\setminus\mathcal{H}_{r-1}
	\label{eq:candidate_satellite_set}
\end{equation}
denote the replacement-candidate satellite set at round $r$. Network planning then compares the networking consequences of the feasible launch choices formed from $\mathcal{C}_r$.
The satellites in $\mathcal{C}_r$ are further organized according to the adopted launch-grouping rule. The initial BDS-3 deployment used two-satellite missions for same-plane MEO satellites and single-satellite missions for IGSO and GEO satellites~\cite{20,21}. The same launch pattern is adopted as the default replacement mode in this study. 
Let $\mathcal{A}_r(\mathcal{C}_r)$ denote the corresponding legal launch-action set. Each $a_r\in\mathcal{A}_r(\mathcal{C}_r)$ contains the satellites deployed in round $r$, and the deployed state evolves as
\begin{equation}
	\mathcal{H}_r
	=
	\mathcal{H}_{r-1}\cup a_r,
	\qquad r=1,2,\ldots,R .
	\label{eq:state_update}
\end{equation}

For $r=1,2,\ldots,R-1$, let $\Delta T_r>0$ denote the interval from the completion of round $r$ to the completion of round $r+1$. The state $\mathcal{H}_r$ is the operating constellation state throughout $\Delta T_r$, so the duration of each inter-launch state determines how long its network capability contributes to the progressive replacement process.

\subsection{Hybrid Network Operation, Time Scales, and Resource Constraints}
\label{subsec:hybrid_operation}

\subsubsection{Hybrid constellation state and high-rate ground access}

The laser satellites in $\mathcal{H}_r$ carry LISL capability together with phased-array terminals, while the satellites in $\mathcal{S}\setminus\mathcal{H}_r$ continue to operate through the legacy phased-array terminals. Laser satellites therefore participate in both the laser and microwave networks, and legacy satellites participate in the microwave network. Each intermediate replacement state consequently contains an expanding laser-enabled part and a residual microwave part that remain coupled through phased-array links.

As shown in Fig.~\ref{fig:hybrid_scenario}, the laser-enabled part becomes a effective high-rate backbone only when its LISLs are connected to the ground segment through high-speed satellite--ground links. Ground access serves two central functions. First, telemetry and service traffic, such as BeiDou global short-message data, must eventually be delivered to ground stations. The current BDS-3 microwave satellites carry legacy S-band satellite--ground links for such data return~\cite{1,2,9}. Their transmission capacity is far below the Gbps-class forwarding capability of the LISL backbone. As progressive replacement expands the laser backbone, the ground-return segment therefore requires high-rate satellite-ground links with a capacity commensurate with the traffic aggregated by the laser backbone. Second, the high-precision timing capability of the laser network relies on ground-referenced clock synchronization and subsequent dissemination of timing information through the laser backbone.

Accordingly, the replacement architecture should equip a subset of laser satellites with additional high-rate satellite--ground terminals. These feeder-capable laser satellites act as ground gateways for the laser backbone.
Although satellite-to-ground laser transmission is a natural high-rate option, its performance is highly sensitive to atmospheric and weather conditions~\cite{22}. A more practical solution is therefore to equip part of the laser satellites with Ka-band large-aperture reflector antennas so as to construct high-rate satellite-to-ground links whose bandwidth can be matched with that of the inter-satellite laser backbone~\cite{23}. These links provide both the data-return path to the ground segment and the physical basis for satellite-to-ground time synchronization.

We use the term \emph{high-rate satellite--ground link resources} for the resources required to establish these links. Their availability is jointly determined by satellite-side high-rate-link capability, satellite--ground visibility, and concurrent-access capacity at the ground segment. A connected set of deployed laser satellites that reaches one or more active high-rate satellite--ground links is termed a \emph{ground-connected laser subnetwork}. This subnetwork represents the portion of the deployed laser capability that operates as a high-rate inter-satellite backbone with effective access to the ground system.

\subsubsection{Visibility abstraction for laser and microwave links}

Satellite motion continuously changes the geometry between node pairs. Two satellites are regarded as mutually visible when their line of sight is free from celestial-body blockage and both satellites lie within the corresponding beam-coverage limits. Visibility is the physical prerequisite for establishing a link.

The continuously varying visibility process is discretized with a finite-state-automaton (FSA) abstraction~\cite{16}. The considered period is partitioned into equal-length FSA states. A node pair is treated as visible in one state when the corresponding visibility condition is satisfied throughout that entire state. Each FSA state therefore provides a static visibility snapshot for topology planning.

Laser and microwave links use different FSA durations because their link-switching mechanisms operate on different time scales. LISL reconfiguration requires sophisticated PAT. Each topology change triggers a new PAT process, and frequent reconfiguration consumes a substantial fraction of the available link time. Laser topology planning therefore uses a relatively long FSA duration, typically on the order of one hour~\cite{10}.

Phased-array microwave links support rapid beam switching among visible objects and are used to collect diversified inter-satellite ranging. With the FSA definition above, a longer state requires a pair to remain continuously visible for a longer interval before that pair is admitted into the state visibility graph. Increasing the microwave FSA duration therefore reduces the effective set of visible partners and the available measurement-contact opportunities. Microwave topology planning consequently uses a much shorter FSA duration, typically on the order of several minutes~\cite{13,14}.

\begin{figure}[t]
	\centering
	\includegraphics[width=\linewidth]{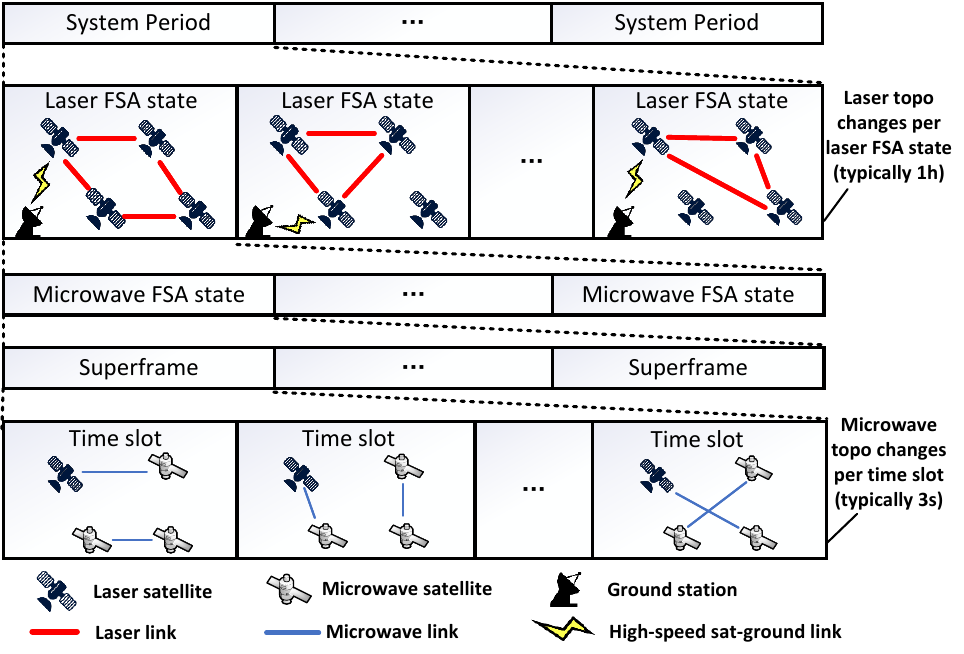}
	\caption{Hierarchical temporal and topology abstraction for the laser-microwave hybrid constellation. Laser topology remains fixed within each FSA state, while microwave visibility uses shorter FSA states with links scheduled via superframes and activated per slot.}
	\label{fig:topology_model}
\end{figure}

Let $\mathcal{K}=\{1,2,\ldots,K\}$ denote the laser FSA states over the visibility-evaluation horizon, and let $\mathcal{Q}_k$ denote the microwave FSA states temporally contained in laser state $k$. Fig.~\ref{fig:topology_model} summarizes this nested temporal abstraction and the corresponding topology-update scales.

\subsubsection{Hierarchical topology evolution}

Within each laser FSA state, the laser topology is fixed. The deployed laser satellites connected by active LISLs therefore form a static graph over that state. Since a laser satellite can carry multiple laser terminals, a connected laser subnetwork can provide simultaneous multi-hop paths and operate as a continuously available high-rate backbone. End-to-end delay inside this backbone is consequently dominated by physical propagation and forwarding delay.

The microwave topology evolves on finer time scales. Each microwave FSA state is divided into a sequence of \emph{superframes}, typically of 1 minute, and each superframe is further divided into \emph{time slots}, typically of 3 seconds. A superframe is the planning unit of the microwave topology scheduler: the link schedule for all slots in one superframe is generated jointly. A time slot is the basic microwave link-activation unit. Under the considered phased-array terminal constraint, one satellite communicates with at most one target in each slot~\cite{1,2,25}. Multi-hop microwave delivery is therefore completed through a sequence of scheduled contacts across multiple slots, producing a store-carry-forward operating regime~\cite{17}. Its dominant delay component is the waiting time for subsequent scheduled contacts, which is quantized by the slot duration. In contrast, millisecond level physical propagation delay is secondary

The two topology layers therefore provide complementary functions during replacement. The laser layer supplies a persistent high-rate backbone within each laser FSA state, while the microwave layer rapidly changes contact partners across slots to support measurement diversity and to connect residual microwave satellites with the evolving laser backbone.

\subsubsection{Network resource constraints}

The feasible laser-network realizations are shaped by three resource conditions. First, each satellite carries a finite number of laser terminals, which limits the number of simultaneously active LISLs incident to that satellite. Second, every active high-rate satellite--ground link consumes satellite-side and ground-side resources, so the number and placement of available ground outlets are constrained by the high-rate satellite--ground link resources defined above.
Third, expansion of the ground-connected laser subnetwork increases the aggregate traffic that can be collected and forwarded by the laser backbone. A larger backbone therefore requires greater aggregate ground-return capacity. The ground-return capacity is scaled with the laser-backbone size through a \emph{feeder requirement}: as the ground-connected laser subnetwork crosses prescribed size levels, a corresponding minimum number of high-rate satellite--ground links must be active. This coupling keeps the growth of the satellite--ground return segment aligned with the traffic-carrying capability of the expanding LISL backbone. State-dependent visibility together with these resource conditions defines the feasible network-realization space of a deployed set.

\subsection{Networking Potential of an Intermediate Deployment State}
\label{subsec:network_potential}

A deployed set $\mathcal{H}\subseteq\mathcal{S}$ specifies which constellation positions already possess laser capability. The network that can be formed from these satellites also depends on the current visibility state and the resource conditions above. Sequence planning therefore requires a state value that captures the best networking capability attainable by the deployed set under a common physical model. We call this state value the \emph{networking potential} of $\mathcal{H}$.

\subsubsection{Maximum attainable laser-network scale}

The first objective is to bring as many deployed laser satellites as possible into the ground-connected laser subnetwork. Deployed laser satellites incorporated into this subnetwork can immediately exploit the high-rate backbone, while deployed laser satellites outside the subnetwork continue to operate through their phased-array microwave terminals during that state. Enlarging the ground-connected laser subnetwork therefore converts a larger fraction of the deployed laser payloads into effective system-level backbone capability.

A larger ground-connected laser subnetwork provides three main benefits. First, every satellite inside the subnetwork can forward telemetry and service traffic through continuously available LISL paths toward a high-rate ground outlet. This operating mode replaces multiple slot-level microwave waiting stages with laser forwarding dominated by propagation and forwarding delay. Second, the connected laser satellites form a high-precision timing and ranging domain. Laser links provide substantially higher synchronization accuracy and continuous high-quality inter-satellite measurements, so enlarging this domain increases the number of satellites directly benefiting from these capabilities. Third, a larger laser backbone enlarges the part of the constellation that can support high-rate service forwarding. For example, once a short-message user accesses a satellite inside the laser backbone, the subsequent message can be forwarded through high-rate LISLs toward the ground segment with low additional network delay~\cite{18}.

For laser FSA state $k$, let $\Omega_k(\mathcal{H})$ denote the set of all network realizations supported by $\mathcal{H}$ that satisfy the visibility and resource conditions above. For a realization $\omega\in\Omega_k(\mathcal{H})$, let $\mathcal{B}_k(\omega)\subseteq\mathcal{H}$ denote the deployed laser satellites included in its ground-connected laser subnetwork. The maximum attainable subnetwork size is
\begin{equation}
	S_k^{\star}(\mathcal{H})
	=
	\max_{\omega\in\Omega_k(\mathcal{H})}
	\left|\mathcal{B}_k(\omega)\right|.
	\label{eq:primary_potential}
\end{equation}
Thus, $S_k^{\star}(\mathcal{H})$ measures the largest portion of the deployed laser capability that can be organized into an effective ground-connected backbone in state $k$.

\subsubsection{Visibility support to the residual microwave segment}

The residual microwave satellites continue to operate through phased-array links and can use the ground-connected laser subnetwork as a high-rate relay domain. When a residual microwave satellite is visible to a backbone satellite in a microwave FSA state, one scheduled microwave link is sufficient to enter the laser backbone. After this one-hop access, the data follow the continuously available laser path toward a high-rate ground outlet during the laser FSA state. A larger set of such visibility opportunities therefore provides more rapid-backbone-access opportunities and shortens the microwave store-carry-forward process preceding ground return.

The same contact pattern also benefits timing synchronization. The laser backbone forms a high-precision timing domain referenced to the ground through its high-rate satellite--ground links. A residual microwave satellite that reaches this domain in one microwave hop obtains timing information through a shorter transfer chain. Synchronization error accumulates along timing-transfer hops, so richer direct visibility from residual microwave satellites to the laser backbone shortens microwave timing-transfer chains and improves time-transfer quality~\cite{19}.

For microwave FSA state $q$, define
\begin{equation}
	\delta_{im}^{(q)}
	=
	\begin{cases}
		1, & \text{if satellites $i$ and $m$ are visible in state $q$,}\\
		0, & \text{otherwise,}
	\end{cases}
	\label{eq:microwave_visibility_indicator}
\end{equation}
where $i\in\mathcal{H}$ is a deployed laser satellite and
$m\in\mathcal{S}\setminus\mathcal{H}$ is a residual microwave satellite.
For realization $\omega$ in laser state $k$, define the accumulated microwave-visibility support as
\begin{equation}
	\begin{aligned}
		\Gamma_k(\omega,\mathcal{H})
		&=
		\sum_{q\in\mathcal{Q}_k}
		\sum_{i\in\mathcal{B}_k(\omega)}
		\sum_{m\in\mathcal{S}\setminus\mathcal{H}}
		\delta_{im}^{(q)} .
	\end{aligned}
	\label{eq:microwave_support_realization}
\end{equation}
A larger $\Gamma_k(\omega,\mathcal{H})$ represents more one-hop access opportunities from the residual microwave segment to the laser backbone over the microwave states contained in laser state $k$.

\subsubsection{Hierarchical networking potential definition}

The existing BDS-3 microwave network is already a complete operational system. Its microwave inter-satellite links and legacy S-band satellite--ground links independently support the established RNSS, inter-satellite measurement, telemetry, and timing functions throughout the replacement process. Progressive replacement adds LISLs to introduce higher-precision timing and ranging together with a high-rate inter-satellite backbone. The ground-connected laser-subnetwork size therefore characterizes how much of this new laser capability has become an effective network, while microwave visibility support characterizes the additional benefit that the new backbone provides to the residual microwave segment. This functional relation gives the two quantities a strict hierarchy: laser-network formation is optimized first, and microwave visibility support refines the choice among realizations with the same maximum laser-network scale.

Accordingly, among the realizations that attain $S_k^{\star}(\mathcal{H})$, the maximum attainable microwave-visibility support is
\begin{equation}
	\begin{aligned}
		C_k^{\star}(\mathcal{H})
		&=
		\max_{\substack{
				\omega\in\Omega_k(\mathcal{H})\\
				|\mathcal{B}_k(\omega)|
				=S_k^{\star}(\mathcal{H})
		}}
		\Gamma_k(\omega,\mathcal{H}) .
	\end{aligned}
	\label{eq:secondary_potential}
\end{equation}

Aggregating the FSA-level optima over the visibility-evaluation horizon gives
\begin{equation}
	\begin{aligned}
		P(\mathcal{H})
		&=
		\sum_{k\in\mathcal{K}}
		S_k^{\star}(\mathcal{H}),\\
		Q(\mathcal{H})
		&=
		\sum_{k\in\mathcal{K}}
		C_k^{\star}(\mathcal{H}).
	\end{aligned}
	\label{eq:state_potential_components}
\end{equation}
The networking potential of deployed state $\mathcal{H}$ is represented by
\begin{equation}
	g^{\star}(\mathcal{H})
	=
	\big(P(\mathcal{H}),Q(\mathcal{H})\big),
	\label{eq:state_network_potential}
\end{equation}
with lexicographic comparison: $P(\mathcal{H})$ is compared first, and $Q(\mathcal{H})$ refines the comparison when the first component is equal. Since both components are obtained over the same feasible-realization space, $g^{\star}(\mathcal{H})$ provides a common measure of the maximum networking capability attainable by the deployed set.

\subsection{Round-Wise and Full-Sequence Planning Formulations}
\label{subsec:planning_formulations}

The networking potential supports two complementary modes of network-aware replacement planning. Round-wise planning can be invoked at any launch round where several engineering-feasible choices remain, providing network information for the current decision while preserving the surrounding engineering plan. Full-sequence planning uses complete future launch information and all inter-launch durations to optimize the entire trajectory of intermediate states. The two formulations therefore provide, respectively, an engineering-adaptable rolling decision mode and a full-horizon network-performance reference.

\subsubsection{Round-wise planning}

At round $r$, the prevailing engineering conditions provide $\mathcal{C}_r$, and the launch-grouping rule generates $\mathcal{A}_r(\mathcal{C}_r)$. The round-wise decision selects the legal action that gives the largest networking potential to the immediate post-launch state:
\begin{equation}
	a_r^{\star}
	\in
	\operatorname*{arg\,max}_{a\in\mathcal{A}_r(\mathcal{C}_r)}^{\mathrm{lex}}
	g^{\star}\!\left(\mathcal{H}_{r-1}\cup a\right).
	\label{eq:roundwise_planning}
\end{equation}
After the selected action is executed, \eqref{eq:state_update} gives the realized deployed state. At the next launch opportunity, the current engineering conditions define a new candidate set $\mathcal{C}_{r+1}$, and the same decision process is repeated. This formulation can be applied throughout the campaign or at selected launch rounds where network differentiation among multiple engineering-feasible choices is useful.

\subsubsection{Full-sequence planning}

A complete campaign plan specifies the admissible launch choices over the entire replacement horizon. Let
\begin{equation}
	\mathbf{x}
	=
	(a_1,a_2,\ldots,a_R)
	\label{eq:full_sequence_definition}
\end{equation}
denote a complete legal launch sequence, and let $\mathcal{X}$ denote the set of all complete sequences satisfying the launch-grouping rules. The deployed state after the first $r$ actions is
\begin{equation}
	\mathcal{H}_r(\mathbf{x})
	=
	\mathcal{H}_0
	\cup
	\bigcup_{\ell=1}^{r} a_{\ell}.
	\label{eq:sequence_prefix_state}
\end{equation}

The sequence-comparison horizon covers the progressive transition from the first completed launch to the final launch. During this horizon, state $\mathcal{H}_r(\mathbf{x})$ contributes for duration $\Delta T_r$, $r=1,\ldots,R-1$. The accumulated sequence-level components are
\begin{equation}
	\begin{aligned}
		J_P(\mathbf{x})
		&=
		\sum_{r=1}^{R-1}
		\Delta T_r\,
		P\!\left(\mathcal{H}_r(\mathbf{x})\right),\\
		J_Q(\mathbf{x})
		&=
		\sum_{r=1}^{R-1}
		\Delta T_r\,
		Q\!\left(\mathcal{H}_r(\mathbf{x})\right).
	\end{aligned}
	\label{eq:time_weighted_sequence_value}
\end{equation}
The full-sequence planning problem is
\begin{equation}
	\mathbf{x}^{\star}
	\in
	\operatorname*{arg\,max}_{\mathbf{x}\in\mathcal{X}}^{\mathrm{lex}}
	\big(J_P(\mathbf{x}),J_Q(\mathbf{x})\big).
	\label{eq:full_sequence_planning}
\end{equation}
The optimum $\mathbf{x}^{\star}$ maximizes the duration-weighted networking potential over the complete legal sequence space $\mathcal{X}$ under the fixed campaign information. It therefore defines the theoretical full-horizon network-performance reference for the replacement campaign and provides a benchmark for assessing rolling planning decisions.

\section{SCF-ILP-Based Networking Potential Computation}
\label{sec:potential_computation}

The networking potential $g^{\star}(\mathcal{H})$ provides the common
state-evaluation basis for both round-wise and full-sequence replacement
planning. Since candidate launch actions and complete replacement sequences are
compared through the networking potential of their intermediate deployed states,
its computation directly determines the computational feasibility of
the subsequent planning methods. For a fixed deployed set $\mathcal{H}$, the
networking potential is obtained from the best feasible network realization in
each laser FSA state. The ground-connected laser subnetwork is first expanded
to its maximum attainable size, after which its visibility support to the
residual microwave satellites is maximized at that size. Time-varying
visibility, laser-terminal resources, high-rate satellite--ground link
resources, and feeder requirements jointly define the feasible realization
space. This section formulates the corresponding single-commodity-flow
ILP and develops incremental FSA reoptimization and deployed-set state reuse
to support efficient networking potential computation.

\subsection{SCF-ILP Formulation for Potential Computation}
\label{subsec:scf_ilp_potential}

\subsubsection{FSA support graph and selection variables}

Consider a deployed set $\mathcal{H}\subseteq\mathcal{S}$ and a laser
FSA state $k\in\mathcal{K}$. Let $\mathcal{G}$ denote the set of ground
stations (GSs) that can provide high-rate satellite--ground access. The node
set is
\begin{equation}
	\mathcal{V}(\mathcal{H})
	=
	\mathcal{H}\cup\mathcal{G}.
	\label{eq:potential_node_set}
\end{equation}
Let $\mathcal{E}^{ss}_{k}(\mathcal{H})$ denote the visible candidate
LISLs between deployed laser satellites, and let
$\mathcal{E}^{sg}_{k}(\mathcal{H})$ denote the available candidate
high-rate satellite--ground links. The candidate edge set is
\begin{equation}
	\mathcal{E}_{k}(\mathcal{H})
	=
	\mathcal{E}^{ss}_{k}(\mathcal{H})
	\cup
	\mathcal{E}^{sg}_{k}(\mathcal{H}).
	\label{eq:potential_edge_set}
\end{equation}

For each node $v\in\mathcal{V}(\mathcal{H})$, define
$z_v\in\{0,1\}$. For a deployed laser satellite $i\in\mathcal{H}$,
$z_i=1$ indicates that satellite $i$ belongs to the selected
ground-connected laser subnetwork. For a ground station
$g\in\mathcal{G}$, $z_g=1$ indicates that the ground station
participates in this subnetwork through active high-rate
satellite--ground links. For each candidate edge
$\{u,v\}\in\mathcal{E}_{k}(\mathcal{H})$, define
$y_{uv}\in\{0,1\}$, where $y_{uv}=1$ indicates that the link is
activated. Link activation requires both end nodes to be selected:
\begin{equation}
	y_{uv}\le z_u,\qquad
	y_{uv}\le z_v,
	\quad
	\forall\{u,v\}\in\mathcal{E}_{k}(\mathcal{H}).
	\label{eq:potential_edge_node}
\end{equation}

\subsubsection{Laser-terminal and high-rate satellite--ground link constraints}

The number of simultaneously active LISLs is limited by the available
laser terminals. Let $d_i^{ss}$ denote the number of LISL terminals on
deployed laser satellite $i$. Then
\begin{equation}
	\sum_{\substack{
			\{i,j\}\in\mathcal{E}^{ss}_{k}(\mathcal{H})
	}}
	y_{ij}
	\le d_i^{ss},
	\qquad
	\forall i\in\mathcal{H}.
	\label{eq:potential_lisl_degree}
\end{equation}

A high-rate satellite--ground link occupies access resources at both
ends. Let $d_v^{sg}$ denote the maximum number of simultaneous
high-rate satellite--ground links supported by node
$v\in\mathcal{V}(\mathcal{H})$. For a satellite, $d_v^{sg}$ represents
its onboard high-rate ground-link capability; for a GS, it
represents its concurrent access capacity. The corresponding resource
constraint is
\begin{equation}
	\sum_{\substack{
			\{v,w\}\in\mathcal{E}^{sg}_{k}(\mathcal{H})
	}}
	y_{vw}
	\le d_v^{sg},
	\qquad
	\forall v\in\mathcal{V}(\mathcal{H}).
	\label{eq:potential_sg_degree}
\end{equation}

The feeder requirement couples the size of the ground-connected laser
subnetwork with the number of active high-rate satellite--ground
links. Let $\eta_{\ell}$ denote the $\ell$th subnetwork-size threshold,
$\beta_{\ell}$ the corresponding minimum number of active feeder
links, and $b_{\ell}\in\{0,1\}$ the indicator associated with this
threshold, for $\ell=1,\ldots,L$. The relation is
\begin{equation}
	\begin{aligned}
		\sum_{i\in\mathcal{H}} z_i
		&\ge \eta_{\ell}b_{\ell},\\
		\sum_{i\in\mathcal{H}} z_i
		&\le \eta_{\ell}-1+|\mathcal{H}|b_{\ell},\\
		\sum_{\{i,g\}\in\mathcal{E}^{sg}_{k}(\mathcal{H})}
		y_{ig}
		&\ge \beta_{\ell}b_{\ell},
		\qquad \ell=1,\ldots,L.
	\end{aligned}
	\label{eq:potential_feeder}
\end{equation}
The first two relations identify the reached size thresholds, while
the third provides the corresponding number of high-rate ground
outlets.

\subsubsection{SCF ground-connectivity constraints}

Every selected laser satellite belongs to the ground-connected laser
subnetwork through an activated path to the ground segment. SCF
expresses this network-wide connectivity requirement with linear flow
constraints.

Replace each edge in $\mathcal{E}_{k}(\mathcal{H})$ by two oppositely
directed arcs and denote the resulting arc set by
$\mathcal{D}_{k}(\mathcal{H})$. For each
$(u,v)\in\mathcal{D}_{k}(\mathcal{H})$, define the auxiliary flow
$f_{uv}\ge0$. For each ground station $g\in\mathcal{G}$, define
$\rho_g\in\{0,1\}$ to indicate the auxiliary root and $s_g\ge0$ as the
flow injected by that root. Let
$n_{\mathcal{H}}=|\mathcal{V}(\mathcal{H})|$. The root and flow
constraints are
\begin{equation}
	\sum_{g\in\mathcal{G}}\rho_g=1,
	\qquad
	\rho_g\le z_g,
	\quad \forall g\in\mathcal{G},
	\label{eq:potential_root}
\end{equation}
\begin{equation}
	f_{uv}
	\le
	(n_{\mathcal{H}}-1)y_{uv},
	\quad
	\forall (u,v)\in\mathcal{D}_{k}(\mathcal{H}),
	\label{eq:potential_flow_capacity}
\end{equation}
\begin{equation}
	0\le s_g\le n_{\mathcal{H}}\rho_g,
	\quad
	\forall g\in\mathcal{G},
	\label{eq:potential_source_bound}
\end{equation}
and
\begin{equation}
	\sum_{g\in\mathcal{G}}s_g
	=
	\sum_{v\in\mathcal{V}(\mathcal{H})}z_v.
	\label{eq:potential_source_total}
\end{equation}

Each selected laser satellite consumes one unit of auxiliary flow:
\begin{equation}
	\sum_{\substack{
			(u,i)\in\mathcal{D}_{k}(\mathcal{H})
	}}
	f_{ui}
	-
	\sum_{\substack{
			(i,w)\in\mathcal{D}_{k}(\mathcal{H})
	}}
	f_{iw}
	=
	z_i,
	\quad
	\forall i\in\mathcal{H}.
	\label{eq:potential_flow_satellite}
\end{equation}
For each selected ground station, the injected source flow enters the
same balance:
\begin{equation}
	\begin{aligned}
		&
		\sum_{\substack{
				(u,g)\in\mathcal{D}_{k}(\mathcal{H})
		}}
		f_{ug}
		+s_g
		-
		\sum_{\substack{
				(g,w)\in\mathcal{D}_{k}(\mathcal{H})
		}}
		f_{gw}
		=z_g,\\
		&\hspace{31mm}\forall g\in\mathcal{G}.
	\end{aligned}
	\label{eq:potential_flow_ground}
\end{equation}
The resulting auxiliary flow connects every selected satellite to the
selected ground root through active LISLs and high-rate
satellite--ground links, thereby defining one ground-connected laser
subnetwork.

\subsubsection{Lexicographic potential objective}

The visibility support contributed by deployed laser satellite $i$ to
the residual microwave satellites during laser FSA state $k$ is
\begin{equation}
	c_{i,k}(\mathcal{H})
	=
	\sum_{q\in\mathcal{Q}_k}
	\sum_{m\in\mathcal{S}\setminus\mathcal{H}}
	\delta_{im}^{(q)}.
	\label{eq:potential_secondary_coefficient}
\end{equation}
Thus, a feasible realization contains
$\sum_{i\in\mathcal{H}}z_i$ satellites in its ground-connected laser
subnetwork and provides
$\sum_{i\in\mathcal{H}}c_{i,k}(\mathcal{H})z_i$ units of visibility
support to the residual microwave segment. The SCF-ILP solves
\begin{equation}
	\max_{\mathrm{lex}}
	\left(
	\sum_{i\in\mathcal{H}}z_i,\,
	\sum_{i\in\mathcal{H}}
	c_{i,k}(\mathcal{H})z_i
	\right)
	\label{eq:potential_lex_objective}
\end{equation}
subject to
\eqref{eq:potential_edge_node}--\eqref{eq:potential_flow_ground}.
The two optimal components are
$S_k^{\star}(\mathcal{H})$ and
$C_k^{\star}(\mathcal{H})$, respectively.

Solving the ILP over all $k\in\mathcal{K}$ yields the FSA-level
potential trajectory of $\mathcal{H}$. Aggregating these optimal values
over the evaluation horizon gives $P(\mathcal{H})$, $Q(\mathcal{H})$,
and $g^{\star}(\mathcal{H})$ as defined in
Section~\ref{subsec:network_potential}.

\subsection{Acceleration of Networking Potential Computation}
\label{subsec:potential_acceleration}

Computing the networking potential of one deployed set requires solving the
SCF-ILP over all laser FSA states, while both round-wise and full-sequence
planning repeatedly evaluate many deployed sets. The computational cost of the
state evaluator therefore depends largely on avoiding repeated model
construction and repeated solution of previously encountered states. Two
complementary mechanisms are adopted for this purpose: incremental
reoptimization across consecutive FSA states and deployed-set state reuse.

\subsubsection{Incremental Reoptimization Across FSA States}

For a fixed deployed set $\mathcal{H}$, all laser FSA states share the same
nodes, terminal resources, feeder requirements, and SCF connectivity
constraints. Orbital motion mainly changes the available links and the
microwave visibility-support coefficients. We therefore define the union edge
set over the evaluation horizon as
\begin{equation}
	\mathcal{E}^{\cup}(\mathcal{H})
	=
	\bigcup_{k\in\mathcal{K}}
	\mathcal{E}_{k}(\mathcal{H}),
	\label{eq:potential_union_edges}
\end{equation}
and construct one SCF-ILP model on $\mathcal{E}^{\cup}(\mathcal{H})$.
When the evaluation proceeds from laser FSA state $k$ to $k+1$, only the
availability of changed links and the coefficients
$c_{i,k}(\mathcal{H})$ are updated. The optimal integer solution of the current
state is further supplied as the initial solution of the next state. Since
adjacent FSA states usually retain most of the same network structure, this
reoptimization avoids rebuilding the ILP from scratch and provides a close
starting point for the next solve. Sequentially solving all
$k\in\mathcal{K}$ yields $S_k^{\star}(\mathcal{H})$ and
$C_k^{\star}(\mathcal{H})$, whose aggregation gives
$g^{\star}(\mathcal{H})$.

\subsubsection{Deployed-Set State Reuse}

Different launch sequences may reach the same deployed set $\mathcal{H}$.
Under the common visibility and resource conditions, such states have the same
networking potential and therefore require the same SCF-ILP evaluation. For
each distinct deployed set, we store $P(\mathcal{H})$, $Q(\mathcal{H})$, and
the FSA-level optimal values
$\{S_k^{\star}(\mathcal{H}),C_k^{\star}(\mathcal{H})\}_{k\in\mathcal{K}}$.
When the same deployed set is encountered again, these values are retrieved
directly without resolving the SCF-ILP. The resulting state reuse is shared by
the round-wise and full-sequence planners, reducing repeated
networking potential evaluations during replacement-sequence search.

\subsection{Representative Topology Realization for Operational Evaluation}
\label{subsec:representative_topology}

The networking potential specifies the best attainable network capability of an
intermediate deployed state, whereas operational evaluation further requires a
concrete ground-connected laser topology in each laser FSA state that realizes
this capability. We use a \emph{Dense Laser Topology} (DLT) as the representative
realization. DLT follows the same visibility, laser-terminal,
high-rate satellite--ground link, feeder, and ground-connectivity conditions as
the networking potential model, thereby mapping the state-level potential to a
specific topology for subsequent performance evaluation.

For laser FSA state $k$, DLT first selects a feasible realization that attains
the maximum ground-connected laser-subnetwork size
$S_k^{\star}(\mathcal{H})$. Among all realizations with this maximum size, it
then selects one attaining the maximum microwave-visibility support
$C_k^{\star}(\mathcal{H})$. These two stages preserve the same hierarchical
networking criterion used in replacement-sequence planning.
After fixing these two potential components, DLT further maximizes the number
of active LISLs:
\begin{equation}
	\max
	\sum_{\{i,j\}\in
		\mathcal{E}^{ss}_{k}(\mathcal{H})}
	y_{ij}.
	\label{eq:dlt_dense_objective}
\end{equation}
This final criterion produces a dense backbone realization while preserving the
networking potential of the deployed state. 
DLT is used as the common topology-realization layer in the evaluation so that
all replacement sequences are compared under the same operational rule. The
replacement-sequence framework itself depends on the networking potential of
each deployed state rather than on a unique topology-construction procedure;
other realization methods that preserve the same potential hierarchy and
network-resource conditions can therefore be incorporated into the evaluation
framework.

\section{Replacement Planning Methods}
\label{sec:replacement_planning_methods}

The preceding sections define the networking potential
$g^{\star}(\mathcal{H})$ as a common value for intermediate deployed states
and provide an efficient SCF-ILP-based evaluator for its efficient computation.
Building on this common state value, this section develops two complementary
planning methods for different decision horizons. 
Round-wise planning selects the network-wise best legal action at
the current launch opportunity, whereas full-sequence planning searches
the complete replacement trajectory for a strong
network-performance reference.

\subsection{Priority-Aware Round-Wise Planning}
\label{subsec:roundwise_planning_method}

\subsubsection{Round-Wise Decision Process}

At launch round $r$, the prevailing engineering conditions provide the
replacement-candidate set $\mathcal{C}_r$, from which the launch-grouping rule
generates the legal action set $\mathcal{A}_r(\mathcal{C}_r)$. According to
\eqref{eq:roundwise_planning}, the planner compares the networking potentials
of the successor states $\mathcal{H}_{r-1}\cup a$ and selects the
lexicographically best action. Because all candidate successor states persist
over the same inter-launch interval, no additional duration weighting is
required within the current round.

After the selected action is executed, the realized deployed set becomes the
starting state of the next launch opportunity, where the engineering conditions
may provide a different candidate set. Round-wise planning can therefore be
invoked whenever multiple engineering-feasible choices remain, while preserving
adaptability to the evolving replacement campaign.

\subsubsection{Priority-Aware Exact Search}

A straightforward implementation evaluates
$g^{\star}(\mathcal{H}_{r-1}\cup a)$ for every legal action. Since each
evaluation invokes the SCF-ILP over the complete laser-FSA horizon, exhaustive
action evaluation can become the dominant cost of round-wise planning. The
priority-aware exact search therefore exploits low-cost structural information
to prioritize and terminate exact potential evaluations while preserving the
same round-wise optimum as exhaustive comparison.

As an initial pruning step, each action is examined under an optimistic
satellite-side high-rate ground-link capability bound. If the maximum number of
high-rate satellite--ground links supportable by the feeder-capable satellites
in the successor deployed set is still insufficient to meet the feeder
requirement associated with the full successor-state size, that action cannot
attain the corresponding full-subnetwork upper bound and is discarded before
the subsequent priority search.
\paragraph*{1) Action-class upper bounds}

For action size $h$, define the comparison class
\begin{equation}
	\mathcal{A}_r^{(h)}
	=
	\left\{
	a\in\mathcal{A}_r(\mathcal{C}_r):
	|a|=h
	\right\}.
	\label{eq:roundwise_action_class}
\end{equation}
Every action in $\mathcal{A}_r^{(h)}$ produces a successor state with
$|\mathcal{H}_{r-1}|+h$ deployed laser satellites. Since at most all deployed
laser satellites can enter the ground-connected laser subnetwork in each laser
FSA state, the primary-potential upper bound of this class is
\begin{equation}
	\overline{P}_r^{(h)}
	=
	K\left(|\mathcal{H}_{r-1}|+h\right).
	\label{eq:roundwise_primary_upper_bound}
\end{equation}
The comparison classes are processed in descending
$\overline{P}_r^{(h)}$, so classes with the largest attainable primary
potential are examined first.

\paragraph*{2) Relaxed ground-reachability screening}

Within one comparison class, actions that remain structurally capable of
attaining \eqref{eq:roundwise_primary_upper_bound} are assigned the highest
evaluation priority. For candidate action $a$, a relaxed visibility graph is
constructed for every laser FSA state using the visible candidate LISLs among
the satellites in $\mathcal{H}_{r-1}\cup a$. A deployed satellite is regarded
as ground-reachable if it can reach, through this relaxed graph, a deployed
satellite that supports high-rate ground access and is visible to at least one
ground station in the corresponding FSA state. If all satellites in
$\mathcal{H}_{r-1}\cup a$ are ground-reachable in every laser FSA state, the
action is included in the high-priority subset
$\mathcal{A}_{r,H}^{(h)}$. This relaxed test identifies actions whose
visibility structure remains compatible with full-subnetwork formation before
exact networking potential evaluation.

\paragraph*{3) Secondary-support ordering and early termination}

The actions in $\mathcal{A}_{r,H}^{(h)}$ are ordered by the microwave
visibility support that would be obtained if the complete successor deployed
set formed the ground-connected laser subnetwork throughout the laser-FSA
horizon:
\begin{equation}
	\begin{aligned}
		\widehat{Q}_r(a)
		&=
		\sum_{k\in\mathcal{K}}
		\sum_{q\in\mathcal{Q}_k}
		\sum_{i\in\mathcal{H}_{r-1}\cup a}
		\sum_{m\in\mathcal{S}\setminus
			(\mathcal{H}_{r-1}\cup a)}
		\delta_{im}^{(q)} .
	\end{aligned}
	\label{eq:roundwise_full_support}
\end{equation}
A larger $\widehat{Q}_r(a)$ gives higher evaluation priority. If an exact
potential computation returns
\begin{equation}
	P(\mathcal{H}_{r-1}\cup a)
	=
	\overline{P}_r^{(h)},
	\label{eq:roundwise_full_primary}
\end{equation}
then all deployed laser satellites in the successor state belong to the
ground-connected laser subnetwork throughout the laser-FSA horizon, and the
secondary potential equals $\widehat{Q}_r(a)$. Because the high-priority
actions are visited in descending $\widehat{Q}_r(a)$, the first action
satisfying \eqref{eq:roundwise_full_primary} is the lexicographic optimum of
that comparison class. The search can therefore terminate the current class
immediately.

\paragraph*{4) Exact fallback and inter-class stopping}

If no high-priority action attains the class upper bound, the remaining actions
in $\mathcal{A}_r^{(h)}$ are evaluated by the same exact networking potential
computation, and the best action of the class is compared with the current
round-wise best. The search then proceeds to the next comparison class.
Because the classes are processed in descending primary-potential upper bound,
the search can terminate globally once the current best primary potential
strictly exceeds the upper bounds of all unvisited classes. Consequently, the
priority-aware procedure returns the same round-wise optimum as exhaustive
exact comparison while avoiding unnecessary networking potential evaluations.

Algorithm~\ref{alg:roundwise_exact_search} summarizes the complete
priority-aware round-wise search.
\begin{algorithm}[t]
	\caption{Priority-Aware Exact Round-Wise Planning}
	\label{alg:roundwise_exact_search}
	\begin{algorithmic}[1]
		\Require $\mathcal{H}_{r-1}$, $\mathcal{C}_r$,
		$\mathcal{A}_r(\mathcal{C}_r)$
		\Ensure round-wise action $a_r^{\star}$
		\State Remove actions failing the feeder-capability check
		\State Partition the remaining actions into
		$\mathcal{A}_r^{(h)}$ and order the classes by
		$\overline{P}_r^{(h)}$
		\State Initialize the current best action
		\For{each comparison class $\mathcal{A}_r^{(h)}$}
		\State Construct $\mathcal{A}_{r,H}^{(h)}$ by relaxed ground reachability
		\State Order $\mathcal{A}_{r,H}^{(h)}$ by
		decreasing $\widehat{Q}_r(a)$
		\For{each $a\in\mathcal{A}_{r,H}^{(h)}$}
		\State Compute
		$g^{\star}(\mathcal{H}_{r-1}\cup a)$
		\State Update the current best action
		\If{$P(\mathcal{H}_{r-1}\cup a)
			=\overline{P}_r^{(h)}$}
		\State Stop traversing the current class
		\EndIf
		\EndFor
		\If{the class upper bound has not been attained}
		\State Evaluate the remaining actions in
		$\mathcal{A}_r^{(h)}$ exactly
		\State Update the current best action
		\EndIf
		\If{the current best primary potential strictly exceeds
			all remaining class upper bounds}
		\State \textbf{break}
		\EndIf
		\EndFor
		\State \Return the current best action as $a_r^{\star}$
	\end{algorithmic}
\end{algorithm}

\subsection{Visibility-Guided Full-Sequence Planning}
\label{subsec:full_sequence_planning_method}

\subsubsection{Search Objective and ALNS Framework}

The full-sequence formulation in
\eqref{eq:full_sequence_planning} compares complete replacement sequences by
the duration-weighted lexicographic objective
$\bigl(J_P(\mathbf{x}),J_Q(\mathbf{x})\bigr)$. Solving this formulation requires
determining both the order of launch actions and the legal grouping of
same-plane MEO satellites. Even with fixed launch groups, $R$ actions already
admit $R!$ possible orders; alternative MEO pairings further enlarge the
search space. Exhaustive enumeration is therefore impractical for the complete
replacement campaign.

Accordingly, we adopt adaptive large-neighborhood search (ALNS) for
full-sequence planning. Rather than exploring the sequence space blindly, the
search uses the visibility structure and networking potential deficits of the
intermediate states to identify where sequence modification is most valuable
and which candidate moves should receive priority. The round-wise planner,
applied over the complete remaining legal action set at each round, provides a
feasible network-greedy sequence as the initial solution. ALNS then modifies
the working sequence through relocation/swap and local exact repair, while a
larger visibility-guided shake is invoked when the search stagnates. 

Every generated sequence is evaluated by the exact
$\bigl(J_P,J_Q\bigr)$ objective. Its intermediate deployed states are evaluated
through the networking potential computation in
Section~\ref{sec:potential_computation}, with previously encountered states
directly reused through the acceleration mechanism in
Section~\ref{subsec:potential_acceleration}. The following search components
specify how weak states are localized and how visibility information guides
the generation and selection of candidate sequence modifications.

\subsubsection{Visibility-Guided ALNS Search}

\paragraph*{1) Weak-state localization}

The visibility-guided search first identifies the intermediate replacement
states with the largest remaining primary-potential gaps and then locates the
laser FSA states in which these gaps occur. The resulting weak states guide
neighborhood generation, while candidate sequences are evaluated by the exact
$\bigl(J_P,J_Q\bigr)$ objective.

For intermediate state $\mathcal{H}_r(\mathbf{x})$, let
$\overline{S}_r(\mathbf{x})$ denote a resource-relaxed upper bound on the
ground-connected laser-subnetwork size attainable in any laser FSA state. The
bound is determined by the deployed-set size, high-rate satellite--ground link
capacities, and feeder requirements, without imposing the detailed
state-dependent visibility structure. Hence,
\begin{equation}
	S_k^{\star}\!\left(\mathcal{H}_r(\mathbf{x})\right)
	\le
	\overline{S}_r(\mathbf{x}),
	\qquad \forall k\in\mathcal{K}.
	\label{eq:fsa_primary_upper_bound}
\end{equation}
The primary deficit in laser FSA state $k$ is defined as
\begin{equation}
	d_{r,k}(\mathbf{x})
	=
	\overline{S}_r(\mathbf{x})
	-
	S_k^{\star}\!\left(\mathcal{H}_r(\mathbf{x})\right)
	\ge 0 .
	\label{eq:fsa_primary_deficit}
\end{equation}
The FSA-level deficits are aggregated according to the duration of the
intermediate state:
\begin{equation}
	\begin{aligned}
		D_r(\mathbf{x})
		&=
		\Delta T_r
		\sum_{k\in\mathcal{K}}
		d_{r,k}(\mathbf{x})\\
		&=
		\Delta T_r
		\left[
		K\overline{S}_r(\mathbf{x})
		-
		P\!\left(\mathcal{H}_r(\mathbf{x})\right)
		\right],
	\end{aligned}
	\label{eq:full_sequence_deficit}
\end{equation}
where a larger $D_r(\mathbf{x})$ indicates a larger and more persistent
primary-potential gap. Rounds with larger deficits therefore receive higher
modification priority. Within a selected weak round, $d_{r,k}(\mathbf{x})$
identifies the laser FSA states in which the gap is concentrated. Once the
primary deficits vanish across the sequence, relaxed-connectivity deficit and
residual-microwave support gap are used to locate the remaining secondary
improvement opportunities.

\paragraph*{2) Relocation/swap with visibility guidance}

Relocation moves a later launch action toward a weak round, allowing its
network contribution to affect earlier intermediate states. Swap exchanges an
action near a weak round with a later action while preserving the other launch
positions. Both operations generate alternative prefix-state trajectories
around the identified weak region.

The generated candidates are ranked lexicographically by three structural
criteria. \emph{Critical-component bridging} first favors moves that, in
primary-deficit FSA states, bring more deployed satellites into the relaxed
ground-reachable component and reduce disconnected components.
\emph{Temporal complementarity} then favors structural improvements occurring
in FSA states with larger $d_{r,k}(\mathbf{x})$, so that candidate changes are
aligned with the time-varying locations of the exact primary deficit.
\emph{Position-specific marginal value} finally compares the cumulative
structural gain over the intermediate states changed by the move, including
relaxed ground-reachable size, ground-access opportunities, laser-link
redundancy, and residual-microwave visibility support. This criterion captures
the different trajectory-level value obtained when the same launch action is
placed at different positions.

After ranking, one candidate is selected from the highest-ranked set for exact
full-sequence evaluation by $\bigl(J_P,J_Q\bigr)$. The same visibility-guided
ranking is used to select the larger relocation/swap moves employed by the
shake mechanism.

\paragraph*{3) Local exact repair}

Local exact repair selects a short window around a weak round and optimizes the
local launch sequence exactly. In \emph{order repair}, the launch actions in
the window are fixed and their legal order is optimized. Because all local
orderings complete the same action set, they reach the same deployed state
after the window; only the intermediate states inside the window contribute to
the local comparison.

When the window contains enough same-plane MEO satellites to admit alternative
legal pairings, \emph{MEO regroup repair} jointly optimizes their pairing and
launch order. Candidate windows are prioritized by their accumulated primary
deficit, relaxed-connectivity deficit, and residual-microwave support gap. The
repair window expands gradually as local stagnation increases.

\paragraph*{4) Adaptive acceptance and diversification}

The relocation/swap and local-repair mechanisms are selected with adaptive
weights. A lexicographically better candidate replaces the working sequence.
Equal-objective moves and limited secondary-objective decreases with an
unchanged primary objective are accepted probabilistically to support
neighborhood exploration. When the search remains stagnant, a \textbf{shake} applies
larger visibility-guided relocation/swap moves around weak states and uses the
resulting sequence as a new working point. Search then resumes with the two
normal mechanisms.

Algorithm~\ref{alg:full_sequence_alns} summarizes the procedure.
\begin{algorithm}[t]
	\caption{Visibility-Guided Full-Sequence Planning}
	\label{alg:full_sequence_alns}
	\begin{algorithmic}[1]
		\Require initial round-wise sequence $\mathbf{x}^{0}$
		\Ensure best-found sequence $\mathbf{x}^{\mathrm{best}}$
		\State $\mathbf{x}\gets\mathbf{x}^{0}$,
		$\mathbf{x}^{\mathrm{best}}\gets\mathbf{x}^{0}$
		\While{the search budget remains}
		\State Locate weak intermediate states
		\State Select relocation/swap or local exact repair
		\State Generate one candidate by the selected mechanism
		\State Evaluate and accept the candidate by
		$\bigl(J_P,J_Q\bigr)$
		\State Update $\mathbf{x}^{\mathrm{best}}$ and operator weights
		\If{the search is stagnant}
		\State Apply a visibility-guided relocation/swap shake
		\EndIf
		\EndWhile
		\State \Return $\mathbf{x}^{\mathrm{best}}$
	\end{algorithmic}
\end{algorithm}
The full-sequence planner therefore combines exact sequence evaluation with
problem-informed neighborhood search. Weak states determine where the sequence
should be modified, visibility structure guides position changes, local exact
repair refines short sequence segments and MEO grouping, and shake provides
large-range diversification when local progress slows.


\section{Evaluation}
\label{sec:evaluation}
This section evaluates the proposed replacement-planning framework under
progressively stronger network participation in launch decisions. The resulting
sequences are compared through the operational performance of their
intermediate laser--microwave hybrid constellation states.

\subsection{Simulation Setup}
\label{subsec:simulation_setup}

\subsubsection{Constellation and network settings}

The simulation uses the actual BDS-3 constellation, consisting of 24 MEO,
3 GEO, and 3 IGSO satellites. The launch grouping follows the historical BDS-3
deployment mode: MEO satellites are launched in same-plane pairs, while GEO and
IGSO satellites are launched individually~\cite{20,21}. This grouping yields
18 launch actions and 17 inter-launch operating states.

Each intermediate deployed state is evaluated over a 7-day visibility horizon.
The laser and microwave FSA durations are set to 1~h and 5~min, respectively.
Each microwave FSA state is further divided into 1-min superframes and 3-s
slots. Three ground stations are considered: Sanya
$(18.23^\circ\mathrm{N},109.02^\circ\mathrm{E})$, Weinan
$(34.31^\circ\mathrm{N},109.28^\circ\mathrm{E})$, and Kashi
$(39.47^\circ\mathrm{N},75.989^\circ\mathrm{E})$. The main simulation
parameters are summarized in Table~\ref{tab:sim_setup}.

\begin{table}[t]
	\centering
	\caption{Main Simulation Parameters}
	\label{tab:sim_setup}
	\begin{tabular}{ll}
		\toprule
		Parameter & Value \\
		\midrule
		Constellation & 24 MEO + 3 GEO + 3 IGSO \\
		Launch actions / operating states & 18 / 17 \\
		Evaluation horizon & 7 days \\
		Laser / microwave FSA duration & 1 h / 5 min \\
		Superframe / slot duration & 1 min / 3 s \\
		\makecell[l]{Laser pointing range\\(MEO/GEO/IGSO)}
		& $70^\circ/80^\circ/80^\circ$ \\
		\makecell[l]{Microwave pointing range\\(MEO/GEO/IGSO)}
		& $60^\circ/45^\circ/45^\circ$ \\
		Ground-station (GS) pointing range & $85^\circ$ \\
		Laser terminals per satellite & 3 \\
		\makecell[l]{High-rate ground-link capability\\(MEO/GEO/IGSO)}
		& $0/1/1$ \\
		Concurrent high-rate links per GS & 2 \\
		Feeder levels & $(1,1),(10,2),(20,3)$ \\
		\bottomrule
	\end{tabular}
\end{table}

\begin{figure}[t]
	\centering
	\includegraphics[width=0.8\linewidth]{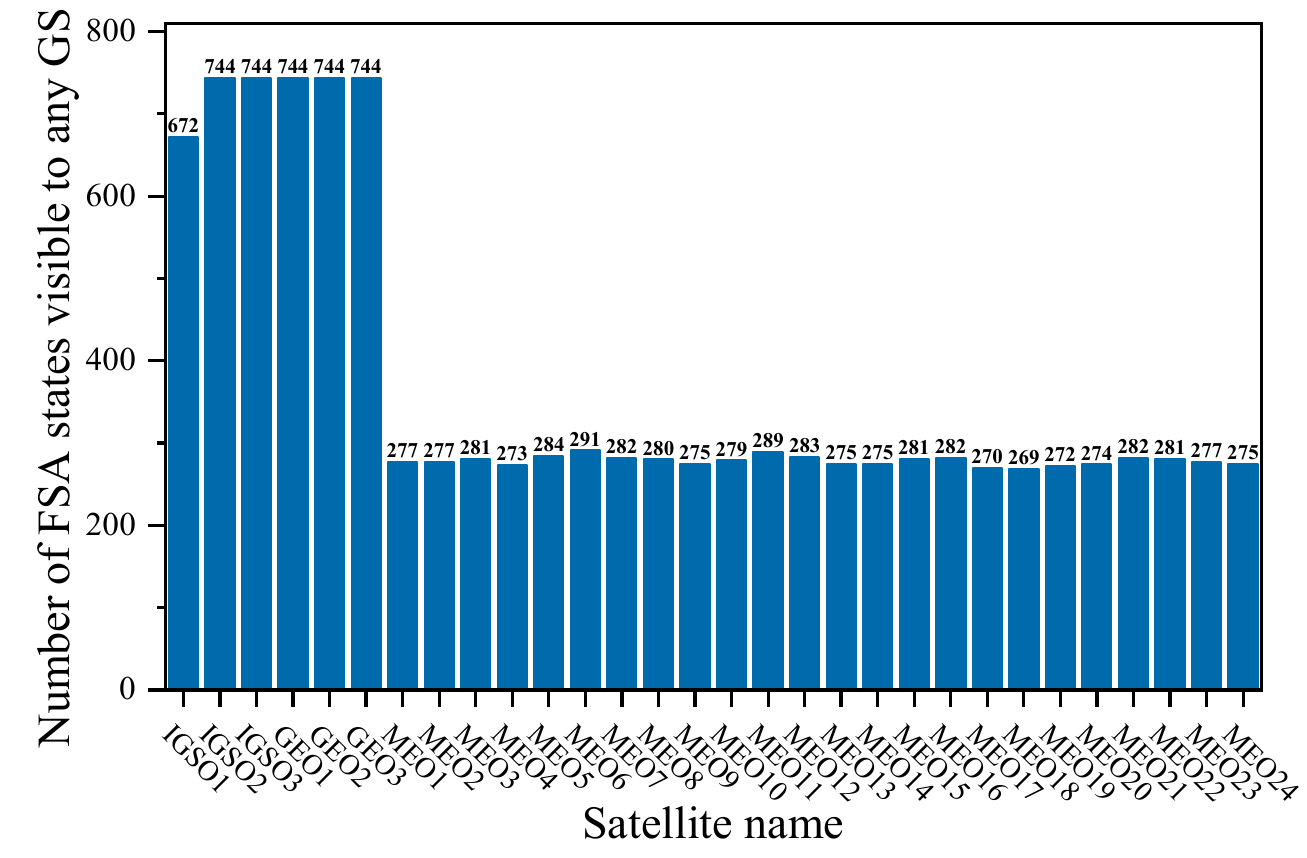}
	\caption{Number of laser FSA states in which each satellite is visible to
		at least one ground station over the 30-day observation window.}
	\label{fig:gs_visibility_fsa_count}
\end{figure}

The high-rate satellite--ground terminals are assigned to GEO and IGSO
satellites according to their ground-access persistence and platform
capability. Fig.~\ref{fig:gs_visibility_fsa_count} shows the number of laser
FSA states in which each satellite is visible to at least one ground station
over a 30-day observation window. The three GEO satellites are visible in all
744 laser FSA states. Two IGSO satellites are also visible in all 744 states,
and the remaining IGSO satellite is visible in 672 states. In comparison, the
24 MEO satellites are visible in 269--291 states. GEO and IGSO satellites
therefore provide substantially more persistent ground-access opportunities.

High-rate satellite--ground terminals also require sufficient payload and power resources. The GEO/IGSO platforms are approximately
5~tons, whereas the MEO platforms are approximately 1~ton~\cite{2}.
Together with their longer ground-access durations, the larger GEO/IGSO
platforms provide more suitable conditions for carrying large-aperture
high-rate ground-link terminals. Accordingly, each GEO/IGSO laser
satellite supports one high-rate satellite--ground link in the simulation,
while the MEO satellites use their legacy S-band ground-access capability. Each
satellite carries three LISL terminals~\cite{24}, and each GS supports at
most two concurrent high-rate links.

The feeder levels are set to
$(1,1)$, $(10,2)$, and $(20,3)$, where the first entry denotes a
ground-connected laser-subnetwork size threshold and the second denotes the
corresponding number of high-rate satellite--ground links. These levels couple
the growth of the laser backbone with the available ground-return capacity. 

For operational evaluation, the representative laser topology is constructed
by DLT in Section~\ref{subsec:representative_topology}. The microwave topology
is scheduled by DFCP~\cite{25} under the single-connection phased-array
constraint. All comparison sequences use the same visibility data, topology
construction rules, and network-resource settings.

\subsubsection{Comparison strategies and network participation}

A practical replacement campaign is jointly influenced by legacy satellite lifetime,
on-orbit condition, laser satellite manufacturing readiness and launch-task scheduling. These factors have different physical meanings and are
difficult to express through a common numerical weight. We therefore
characterize the strength of network assistance by the fraction of launch
rounds in which networking information participates in action selection.
Let $\rho$ denote the network-participation ratio. Four round-wise settings are
considered:
\[
\rho\in\{0,\,0.25,\,0.50,\,0.75\},
\]
corresponding to 0, 4, 9, and 14 network-assisted rounds among the 18 launch
rounds. In a network-assisted round, the round-wise planner selects the
feasible launch action with the largest networking potential.
The remaining engineering-side decisions are represented through randomized
sampling, which statistically captures the combined influence of the other
replacement factors.

For each value of $\rho$, 100 paired samples are generated. The same random
engineering background is shared across different $\rho$ values, so increasing
$\rho$ progressively replaces engineering-driven decisions with network-assisted
decisions under comparable replacement conditions. This design provides an
indirect way to examine how stronger network participation affects the
multi-round replacement process.

The historical BDS-3 deployment order~\cite{20,21} is included as a
lifetime-oriented engineering reference. Under a common design-lifetime
assumption, satellites launched earlier naturally approach the end of their
service life earlier and would therefore tend to enter replacement earlier.
Following the historical deployment order thus provides a simple reference for
a replacement process primarily driven by lifetime progression rather than
network evolution.

For $\rho<1$, networking information affects only part of the launch process,
while the remaining rounds are still determined by engineering-side factors.
The round-wise planner is therefore used to introduce network assistance only
at the selected rounds. 
At $\rho=1$, all launch rounds
are network-driven, so the decision problem changes from selecting locally
preferred actions at individual rounds to determining the complete replacement
sequence from a full-horizon perspective. We therefore use the full-sequence
ALNS planner in Sec~\ref{subsec:full_sequence_planning_method} at this endpoint to search the complete legal sequence space and
obtain a high-quality network-oriented full-horizon reference.

For the simulation, all inter-launch intervals are assumed to have equal
duration, i.e., $\Delta T_1=\cdots=\Delta T_{R-1}$. This simplified setting
removes differences caused solely by launch timing and allows the comparison to
focus on the effect of replacement order and network participation. The
duration-weighted formulation in Section~\ref{sec:system_model}
remains applicable to campaigns with unequal inter-launch intervals.

Table~\ref{tab:comparison_strategies} summarizes the six comparison strategies.
\begin{table}[t]
	\centering
	\caption{Replacement-Sequence Comparison Strategies}
	\label{tab:comparison_strategies}
	\small
	\begin{tabular}{
			p{0.23\linewidth}
			p{0.25\linewidth}
			p{0.40\linewidth}}
		\toprule
		Strategy & Network participation & Role in evaluation \\
		\midrule
		Historical BDS-3
		& None.
		& Lifetime-oriented engineering reference. \\
		
		$\rho=0$
		& 0 of 18 rounds.
		& Engineering-driven reference. \\
		
		$\rho=0.25$
		& 4 of 18 rounds.
		& Low network participation. \\
		
		$\rho=0.50$
		& 9 of 18 rounds.
		& Medium network participation. \\
		
		$\rho=0.75$
		& 14 of 18 rounds.
		& High network participation. \\
		
		ALNS ($\rho=1$)
		& All rounds, full-sequence.
		& Full-horizon network-oriented reference. \\
		\bottomrule
	\end{tabular}
\end{table}
These strategies characterize the gradual transition from engineering-driven
replacement to increasing network assistance and finally to full-horizon
network-oriented planning. All resulting intermediate hybrid constellation
states are evaluated under the same operational model.
The complete historical BDS-3 and best-found ALNS replacement sequences used
in the evaluation are provided in Appendix~\ref{app}.
	
\subsection{Computational Feasibility and Search Efficiency}
\label{subsec:computational_efficiency}

Both round-wise and full-sequence planning repeatedly invoke the
networking potential evaluation of deployed states. The computational cost of
this state evaluator therefore determines the basic efficiency of the overall
planning framework. All experiments are performed using Gurobi 12.0 on a
personal computer with an AMD Ryzen 7 7700X 8-Core processor and 32~GB memory.

For one deployed state, the 7-day evaluation horizon contains 168 laser FSA
states, and the SCF-ILP is reoptimized sequentially over these states. The
average SCF-ILP optimization time is approximately 2.66~ms per laser FSA state.
Including the complete 168-state trajectory and the associated state updates,
one 7-day deployed-state potential evaluation takes 0.496~s on average. This
sub-second state evaluation provides the common computational basis for the two
replacement-planning modes.

For round-wise planning, the priority-aware exact search further reduces the
number of deployed states requiring full potential evaluation. The efficiency
test considers all 18 launch rounds with networking information participating
in each round-wise decision. Across the complete replacement process, 439
engineering-feasible action instances enter the search. Structural screening,
comparison-class ordering, and visibility-support ranking reduce the required
exact networking potential evaluations to 111, corresponding to a 74.7\%
reduction relative to evaluating every engineering-feasible action. The
priority-aware search therefore combines the sub-second state evaluator with a
substantially smaller exact-evaluation set at each launch stage.

Full-sequence planning builds on the same state evaluator and further exploits
deployed-set state reuse. Within a 5h ALNS search budget, 17,508 search
iterations are completed. Repeated intermediate deployed states are retrieved
directly from the shared state cache, yielding a cache-hit ratio of 98.65\%.
The fast state-level SCF-ILP evaluation together with extensive state reuse
supports sustained exploration of complete replacement sequences under the
duration-weighted networking objective.

Table~\ref{tab:computational_efficiency} summarizes the main computational
results.

\begin{table}[t]
	\centering
	\caption{Computational Feasibility and Search Efficiency}
	\label{tab:computational_efficiency}
	\begin{tabular}{ll}
		\toprule
		Metric & Value \\
		\midrule
		Laser FSA states per deployed state & 168 \\
		Average SCF-ILP time per FSA state & 2.66 ms \\
		Average 7-day state-potential evaluation & 0.496 s \\
		\midrule
		Round-wise feasible action instances & 439 \\
		Round-wise exact potential evaluations & 111 \\
		Reduction in exact potential evaluations & 74.7\% \\
		\midrule
		ALNS search budget & 5 h \\
		ALNS iterations & 17,508 \\
		ALNS state-cache hit ratio & 98.65\% \\
		\bottomrule
	\end{tabular}
\end{table}

Taken together, the computational feasibility of the framework is supported by
complementary acceleration mechanisms at multiple levels. Within each deployed
state, incremental reoptimization across consecutive laser FSA states avoids
repeated model construction and reuses neighboring-state solutions. In
round-wise planning, feeder-capability pruning, action-class upper bounds,
relaxed ground-reachability screening, secondary-support ordering, and early
termination substantially reduce the number of deployed states requiring exact
networking potential evaluation. In full-sequence planning, deployed-set state
reuse avoids repeated SCF-ILP solutions for states reached by different
candidate sequences, while the visibility-guided search concentrates sequence
exploration on structurally promising modifications. Together, these mechanisms
enable efficient exact state evaluation and sustained search over both
round-wise actions and complete replacement sequences.
	
\subsection{Network-Performance Results}
\label{subsec:network_performance_results}

\subsubsection{Hybrid-Network Structure Evolution}
\label{subsubsec:hybrid_network_structure}

The structural evolution of the hybrid constellation is characterized from two
attributes: membership in the ground-connected laser subnetwork and direct
visibility to the ground segment. Based on these attributes, every satellite is
assigned to one of four disjoint types, as summarized in
Table~\ref{tab:node_type_definition}.

\begin{table}[t]
	\centering
	\caption{Node-Type Definition in the Hybrid Network}
	\label{tab:node_type_definition}
	\small
	\setlength{\tabcolsep}{3pt}
	\begin{tabular}{
			p{0.08\linewidth}
			p{0.32\linewidth}
			p{0.48\linewidth}}
		\toprule
		Type & Definition & Operational implication \\
		\midrule
		A1
		& In laser subnetwork, GS-invisible
		& Anchor satellite; ground access through the laser backbone and a
		high-rate ground outlet \\
		
		A2
		& In laser subnetwork, GS-visible
		& Anchor satellite; direct legacy S-band access and laser-assisted
		high-rate ground access \\
		
		A3
		& Outside laser subnetwork, GS-visible
		& Anchor satellite; direct ground access through the legacy S-band link \\
		
		A4
		& Outside laser subnetwork, GS-invisible
		& Non-anchor satellite; ground access through microwave contacts with
		A1--A3 satellites \\
		\bottomrule
	\end{tabular}
\end{table}

\begin{figure*}[t]
	\centering
	
	\begin{minipage}[t]{0.3\textwidth}
		\centering
		\includegraphics[width=\linewidth]{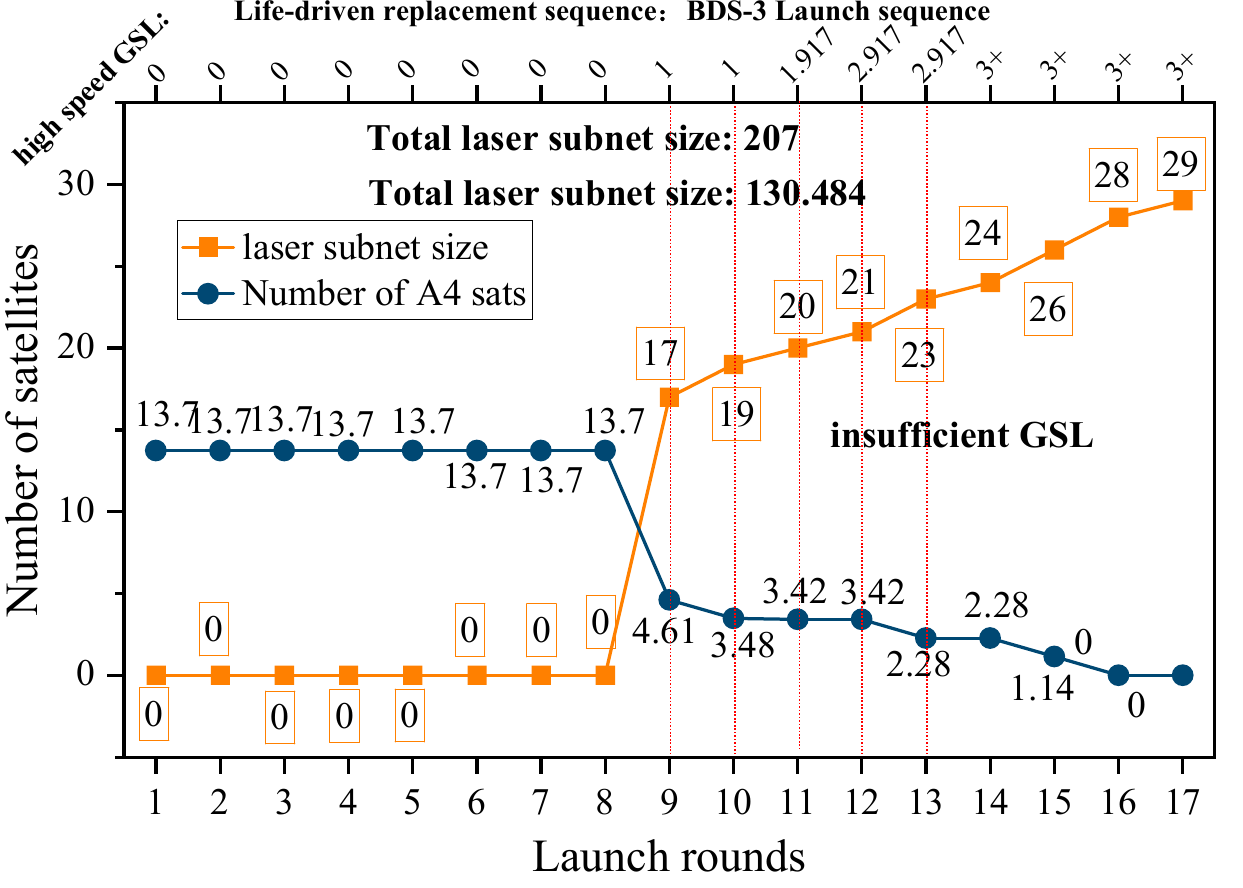}\\[-1mm]
		\textbf{(a)} Historical BDS-3
	\end{minipage}
	\hfill
	\begin{minipage}[t]{0.3\textwidth}
		\centering
		\includegraphics[width=\linewidth]{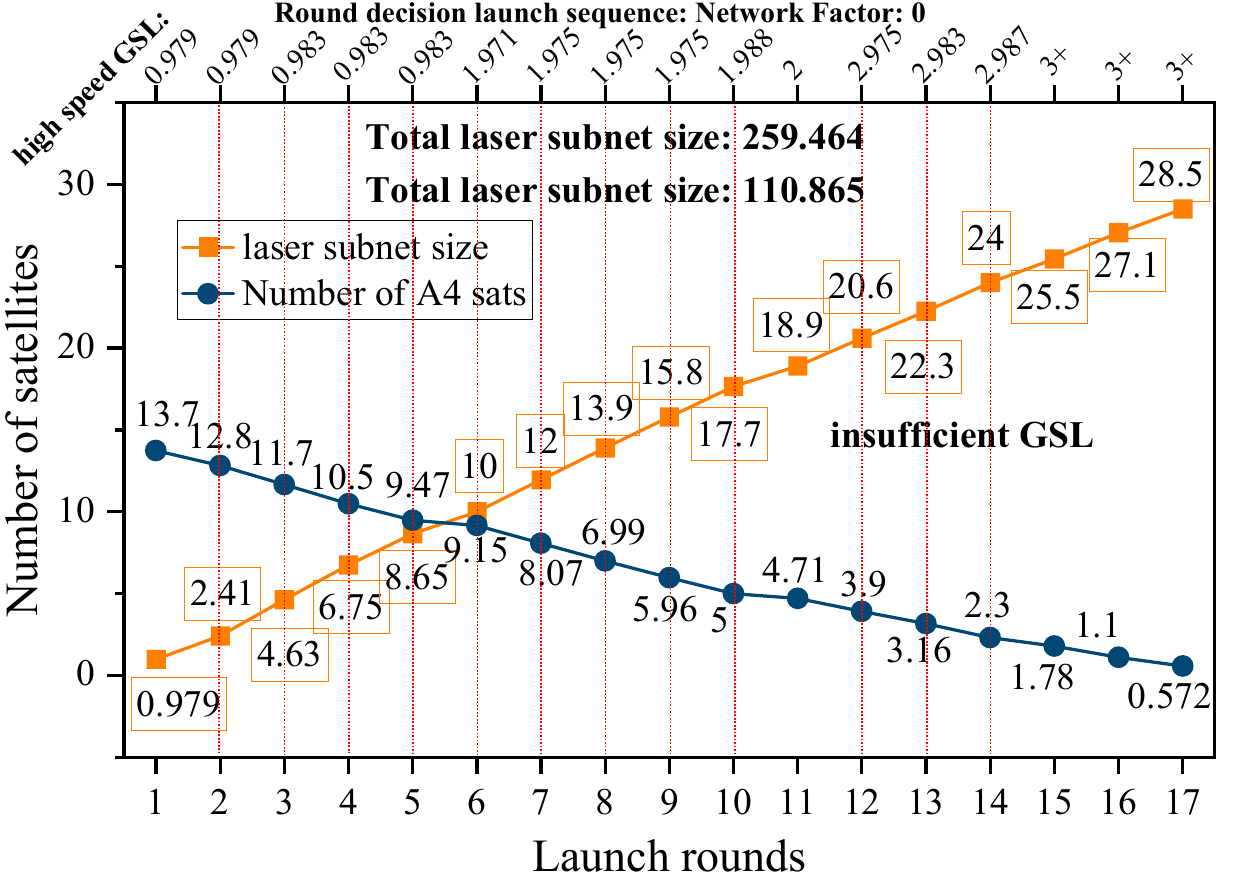}\\[-1mm]
		\textbf{(b)} $\rho=0$
	\end{minipage}
	\hfill
	\begin{minipage}[t]{0.3\textwidth}
		\centering
		\includegraphics[width=\linewidth]{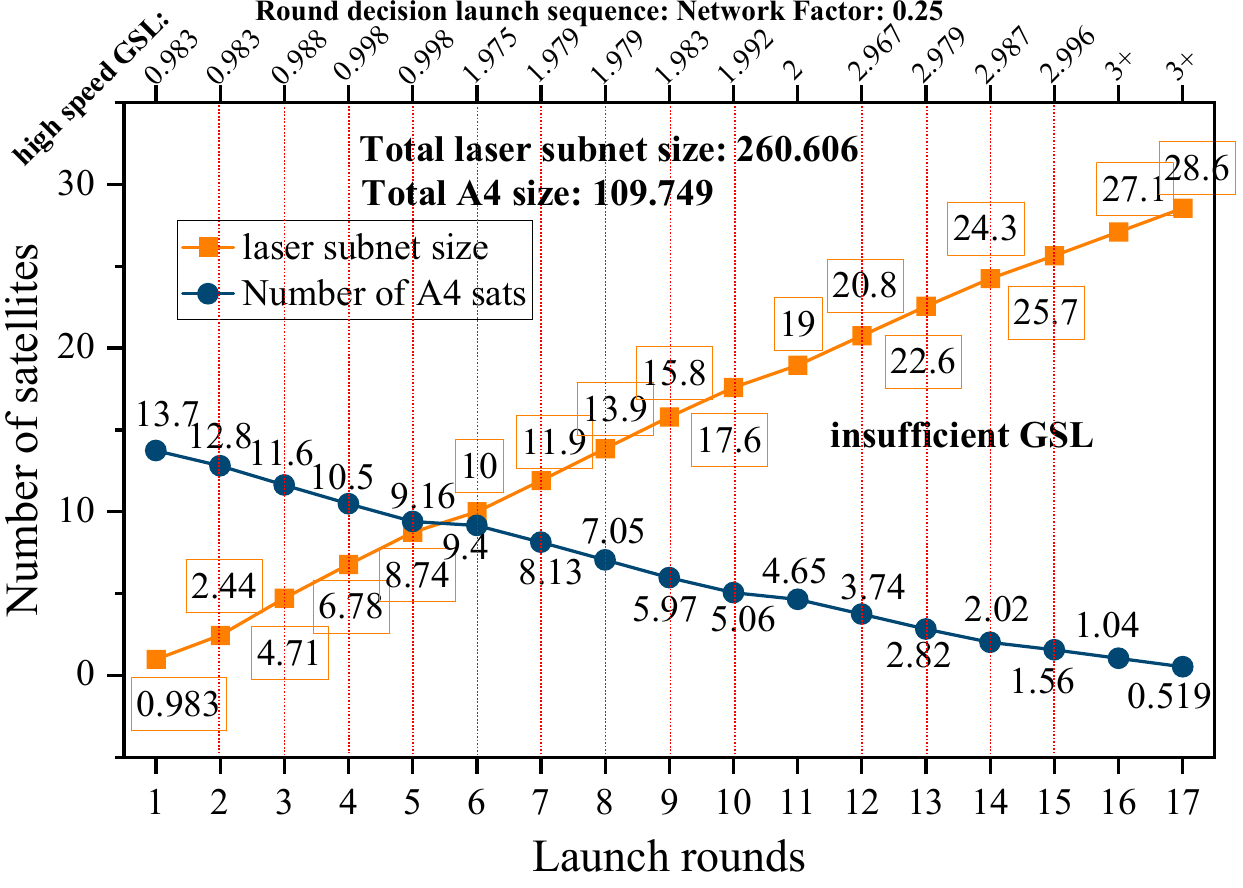}\\[-1mm]
		\textbf{(c)} $\rho=0.25$
	\end{minipage}
	
	\vspace{2mm}
	
	\begin{minipage}[t]{0.3\textwidth}
		\centering
		\includegraphics[width=\linewidth]{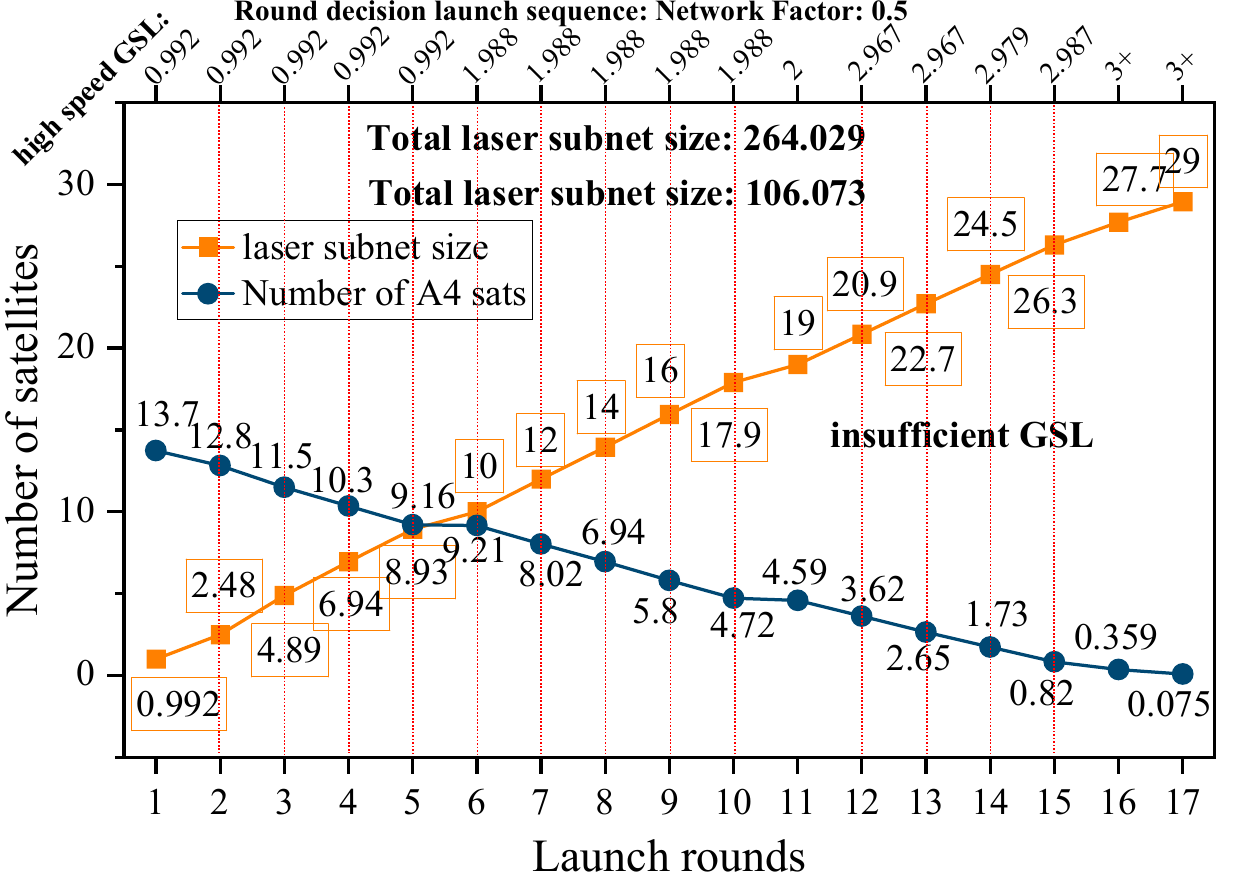}\\[-1mm]
		\textbf{(d)} $\rho=0.50$
	\end{minipage}
	\hfill
	\begin{minipage}[t]{0.3\textwidth}
		\centering
		\includegraphics[width=\linewidth]{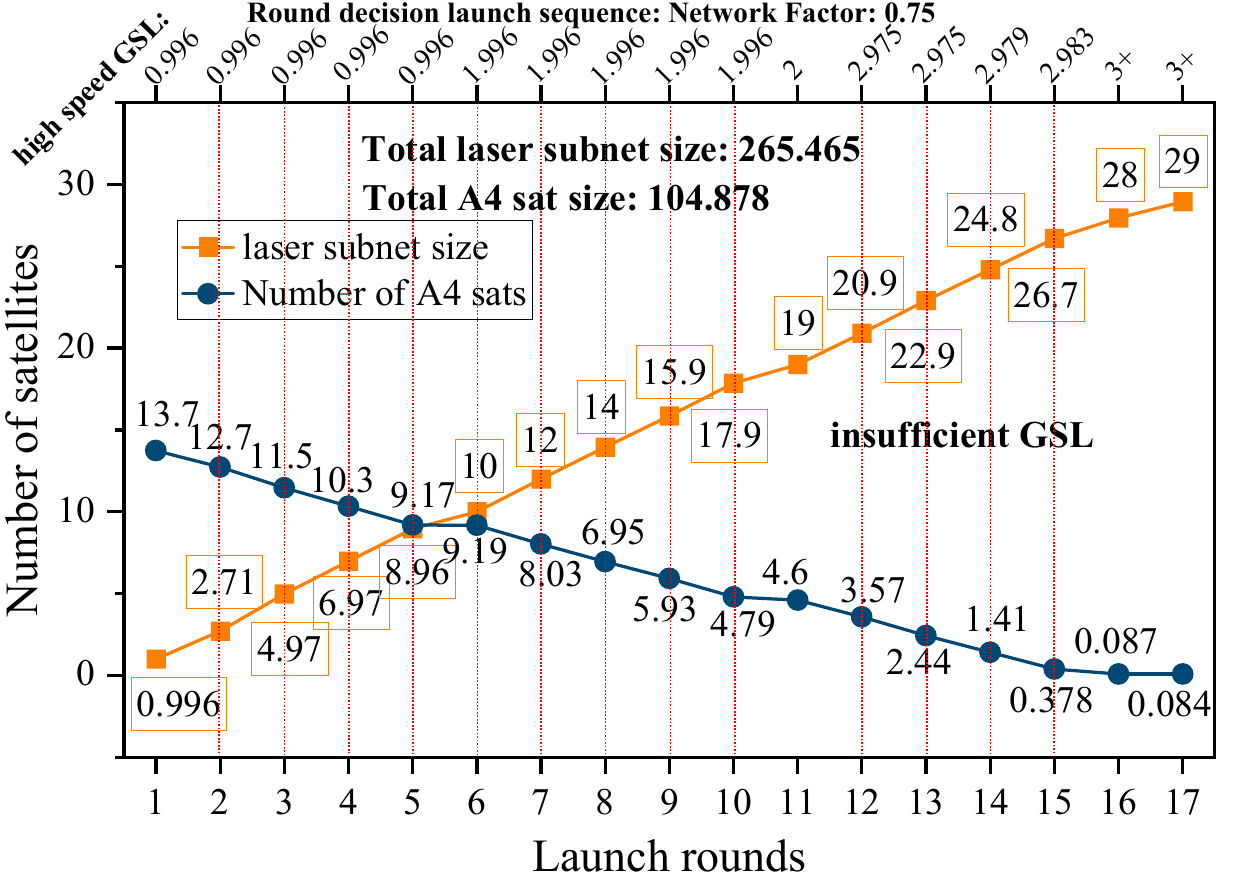}\\[-1mm]
		\textbf{(e)} $\rho=0.75$
	\end{minipage}
	\hfill
	\begin{minipage}[t]{0.3\textwidth}
		\centering
		\includegraphics[width=\linewidth]{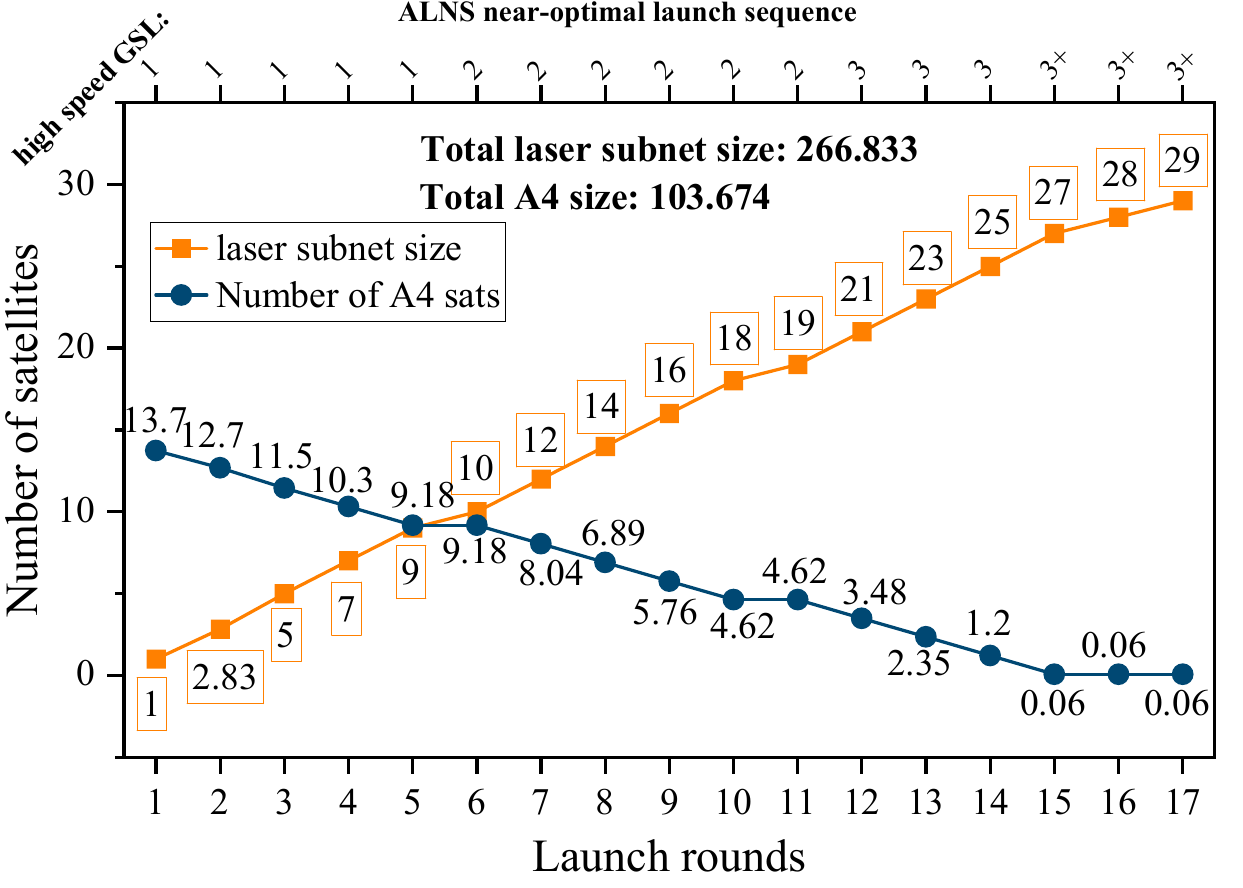}\\[-1mm]
		\textbf{(f)} ALNS
	\end{minipage}
	
	\caption{Evolution of the ground-connected laser subnetwork, A4-satellite
		population, and active high-rate satellite--ground links under the six
		replacement strategies. Red dotted markers denote launch rounds with
		insufficient feeder provisioning relative to the realized laser-subnetwork
		size.}
	\label{fig:laser_subnet_size_2x3}
\end{figure*}
A1 and A2 jointly constitute the ground-connected laser subnetwork. An A1
satellite has no direct ground visibility, but reaches a high-rate ground outlet
through multi-hop LISLs inside the laser backbone. An A2 satellite is both part
of the laser backbone and directly visible to the ground. A3 satellites remain
outside the laser backbone but retain direct ground access through their legacy
S-band links. The union of A1, A2, and A3 forms the anchor set, covering the
satellites that can reach the ground either directly or through the
ground-connected laser backbone. A4 satellites are outside the laser backbone
and are also invisible to the ground; their telemetry return and timing
synchronization therefore rely on intermittent microwave contacts with anchor
satellites. In particular, when a ground-invisible satellite is incorporated
into the laser backbone, it becomes an A1 anchor and gains ground access through
the laser domain. The A1--A4 classification therefore captures both the
expansion of the continuously available high-rate laser domain and the
contraction of the residual satellite population that depends on microwave
store-carry-forward access.

Fig.~\ref{fig:laser_subnet_size_2x3} compares the structural evolution of the
six replacement strategies. The orange curve gives the average size of the
ground-connected laser subnetwork, the blue curve gives the average number of
A4 satellites, and the upper axis gives the average number of active high-rate
satellite--ground links. Red dotted markers identify launch rounds in which the
available high-rate ground links fall below the feeder provisioning target
associated with the current laser-subnetwork size. Fractional values result
from averaging across the evaluated laser FSA states and, for the mixed
round-wise strategies, across the paired samples.

The cumulative results exhibit a consistent improvement as networking
information participates in more launch decisions. The 17-round cumulative
laser-subnetwork size increases from 259.464 at $\rho=0$ to 260.606,
264.029, and 265.465 at $\rho=0.25$, $0.50$, and $0.75$, respectively,
while the cumulative A4 size decreases from 110.865 to 109.749, 106.073,
and 104.878. The full-sequence ALNS result further reaches a cumulative
laser-subnetwork size of 266.833 and an A4 size of 103.674. For comparison,
the historical sequence gives 207.000 and 130.484, respectively. A larger
ground-connected laser subnetwork places more satellites inside the persistent
high-rate forwarding, precise timing, and continuous ranging domain, while a
smaller A4 population directly reduces the number of satellites that depend on
intermittent microwave contacts for ground return and timing access.

The round-wise evolution explains the large gap between the historical
sequence and the network-assisted strategies. All mixed round-wise strategies
and ALNS establish a ground-connected laser subnetwork from the first launch
round and expand it progressively. Under the historical BDS-3 order, the laser
subnetwork remains absent during the first eight rounds and first appears at
Round~9, when the first feeder-capable GEO satellite is introduced. The
historical deployment order places MEO launches ahead of most GEO/IGSO
launches, which is consistent with its original objective of establishing the
BDS-3 RNSS service. Progressive replacement starts from an already
operational microwave constellation, so early introduction of feeder-capable
GEO/IGSO satellites creates greater value by activating the laser backbone
during the intermediate replacement stages. This difference also explains the
much larger early-round A4 population under the historical sequence.

The feeder-link evolution provides a second structural dimension. The
provisioning targets require one, two, and three active high-rate
satellite--ground links when the laser-subnetwork size reaches 1, 10, and 20,
respectively. As $\rho$ increases, the active feeder links follow these
subnetwork-size transitions more closely and the feeder-shortage intervals
become less pronounced. Under ALNS, the feeder-link count follows the
$1\rightarrow2\rightarrow3$ progression together with backbone growth across
the complete replacement process. The mixed strategies exhibit small
state-averaged deviations, such as 0.992 or 1.988, around the corresponding
thresholds. A value below one reflects intermittent loss of the first high-rate
ground outlet over the evaluated states, interrupting the ground-connected
laser-backbone service during those intervals. A value slightly below two
reflects intermittent shortage of the second outlet while one ground path
remains available; the backbone remains ground-connected, while the
ground-return segment provides less capacity than the provisioning target and
can become the bandwidth bottleneck. The historical sequence exhibits the same
coupling issue more prominently: when its first ground-connected laser
subnetwork appears at Round~9, the average backbone already contains 17
satellites but only one high-rate ground link is active.

Overall, increasing network participation improves the intermediate hybrid
states along three coupled dimensions: a larger ground-connected laser
subnetwork, a smaller residual A4 population, and closer matching between
laser-backbone growth and high-rate ground-return capacity. This progression
is consistent with the hierarchical networking potential used in replacement
planning. Its primary component prioritizes the maximum attainable
ground-connected laser-subnetwork size, so stronger network participation
increasingly favors launch actions that incorporate more deployed laser
satellites into the effective backbone. As the backbone expands, more
ground-invisible satellites become A1 anchors and the residual A4 population
correspondingly decreases. 
At the attained backbone scale, the secondary
component further favors visibility support from the laser subnetwork to the
residual microwave segment, while the feeder requirement couples backbone
growth with the required high-rate ground outlets. The observed structural
evolution therefore directly reflects the planning objectives and resource
coupling embedded in the proposed network-potential model.

\subsubsection{Laser-Subnetwork Delay and Link Utilization}
\label{subsubsec}

\begin{figure*}[t]
	\centering
	
	\begin{minipage}[t]{0.3\textwidth}
		\centering
		\includegraphics[width=\linewidth]{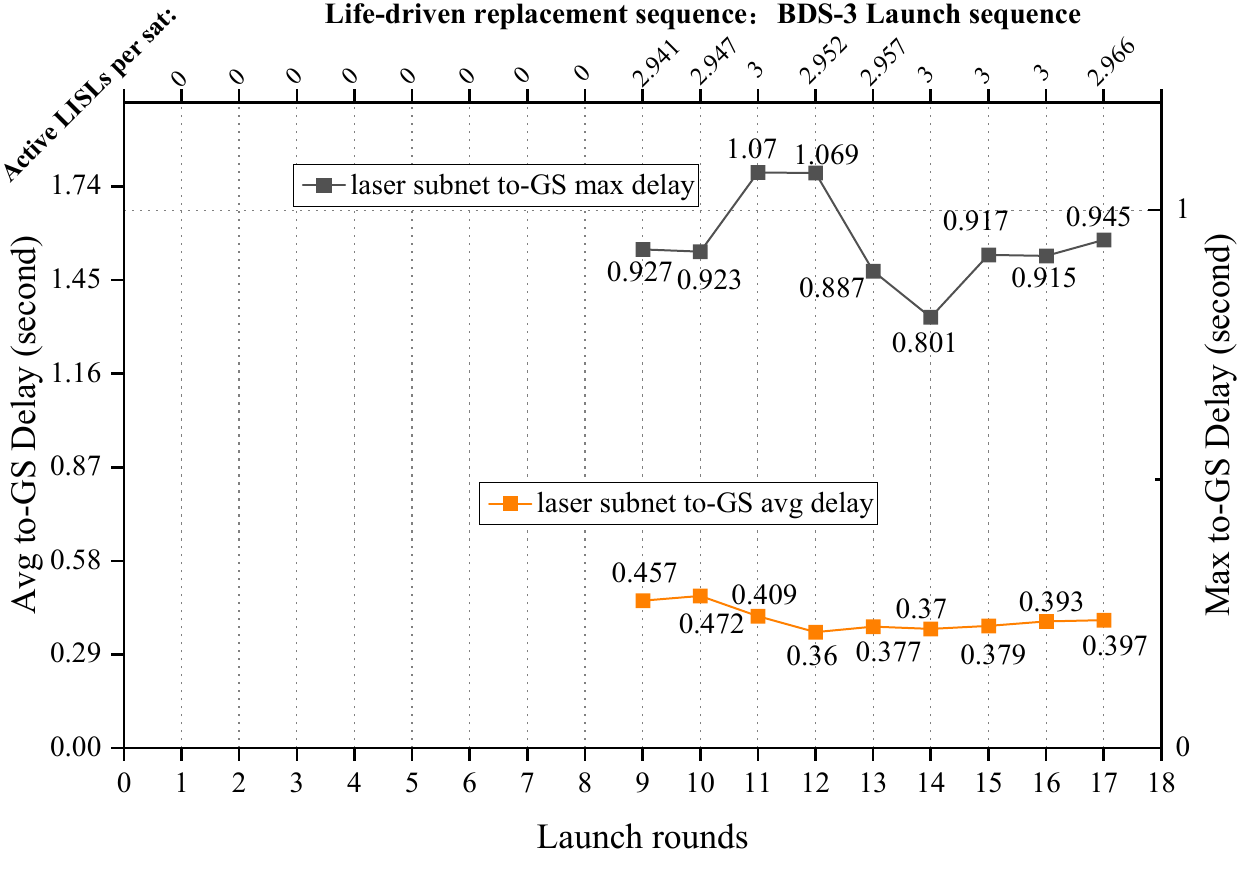}\\[-1mm]
		\textbf{(a)} Historical BDS-3
	\end{minipage}
	\hfill
	\begin{minipage}[t]{0.3\textwidth}
		\centering
		\includegraphics[width=\linewidth]{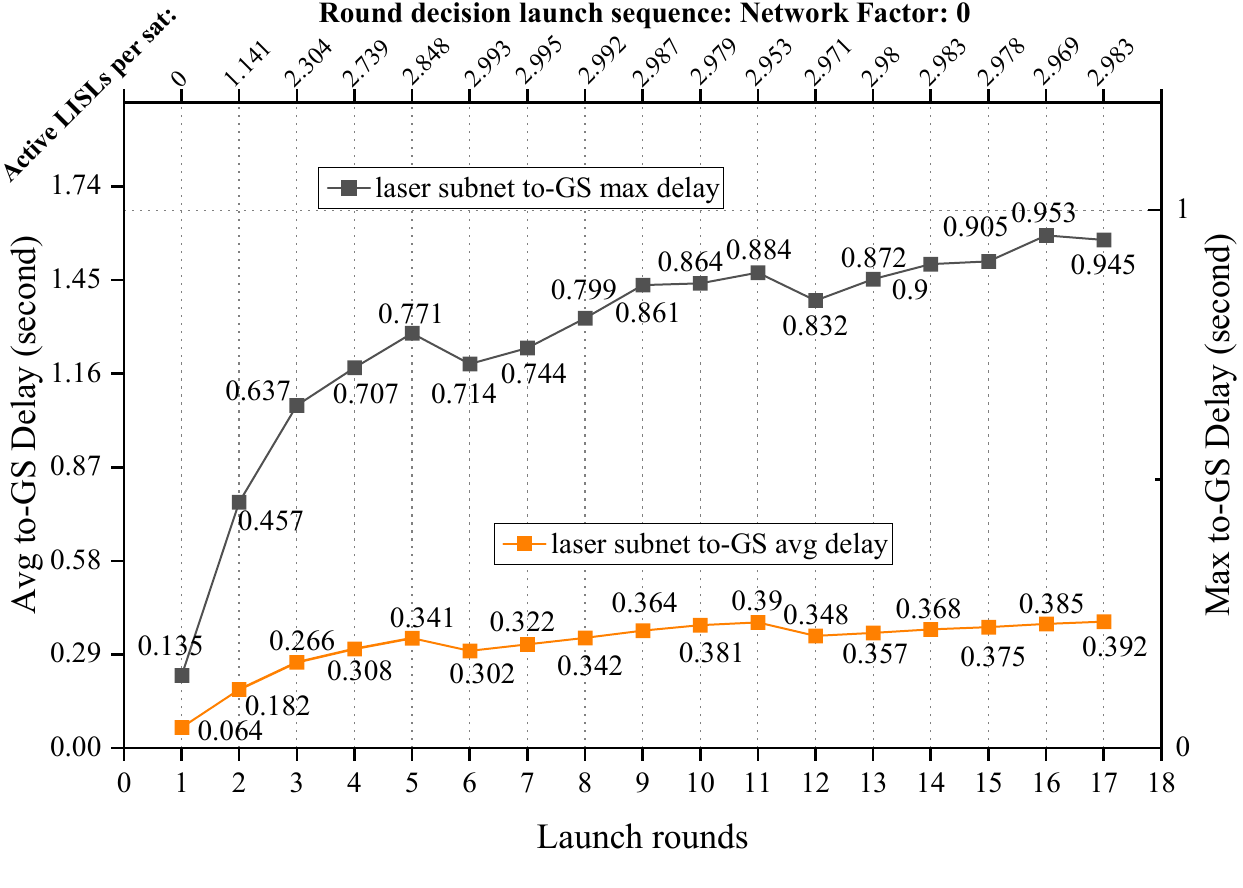}\\[-1mm]
		\textbf{(b)} $\rho=0$
	\end{minipage}
	\hfill
	\begin{minipage}[t]{0.3\textwidth}
		\centering
		\includegraphics[width=\linewidth]{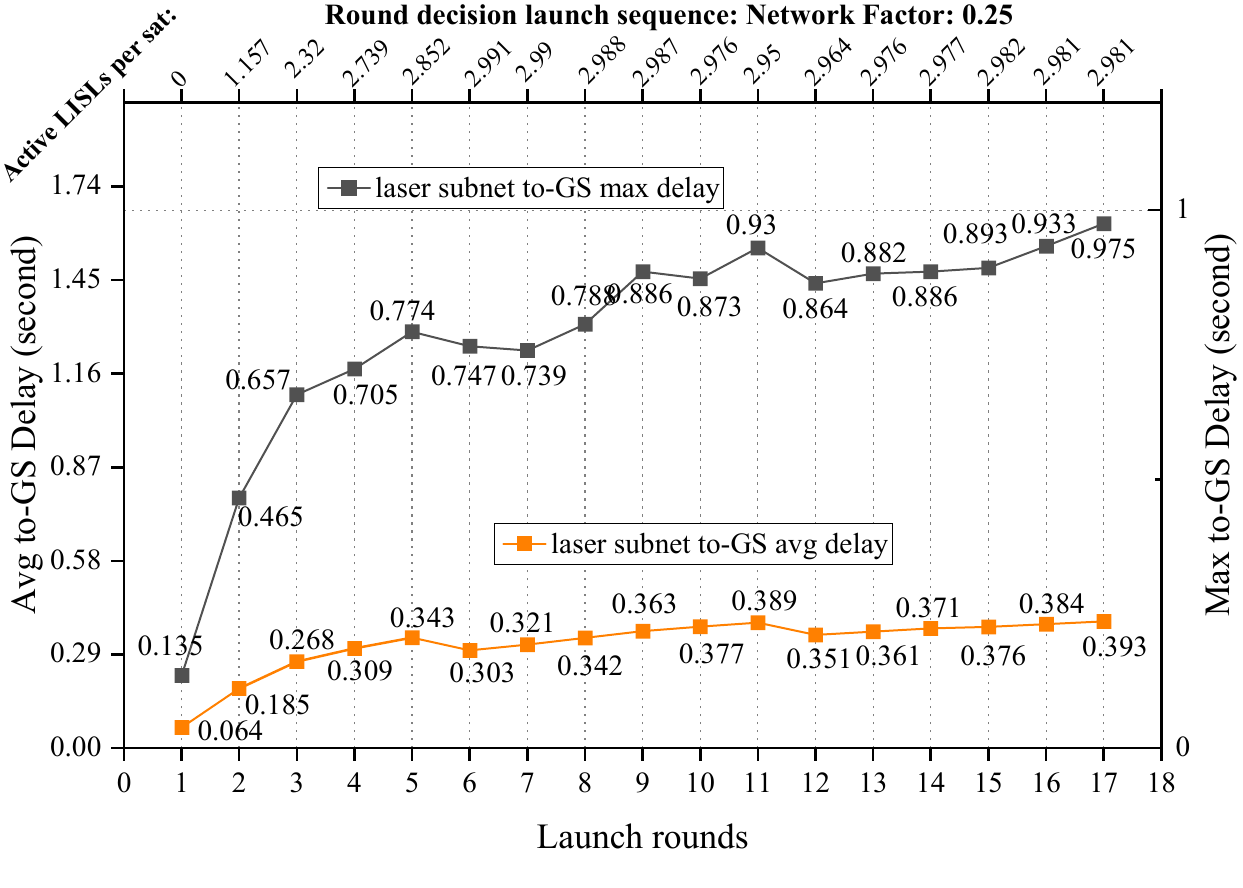}\\[-1mm]
		\textbf{(c)} $\rho=0.25$
	\end{minipage}
	
	\vspace{2mm}
	
	\begin{minipage}[t]{0.3\textwidth}
		\centering
		\includegraphics[width=\linewidth]{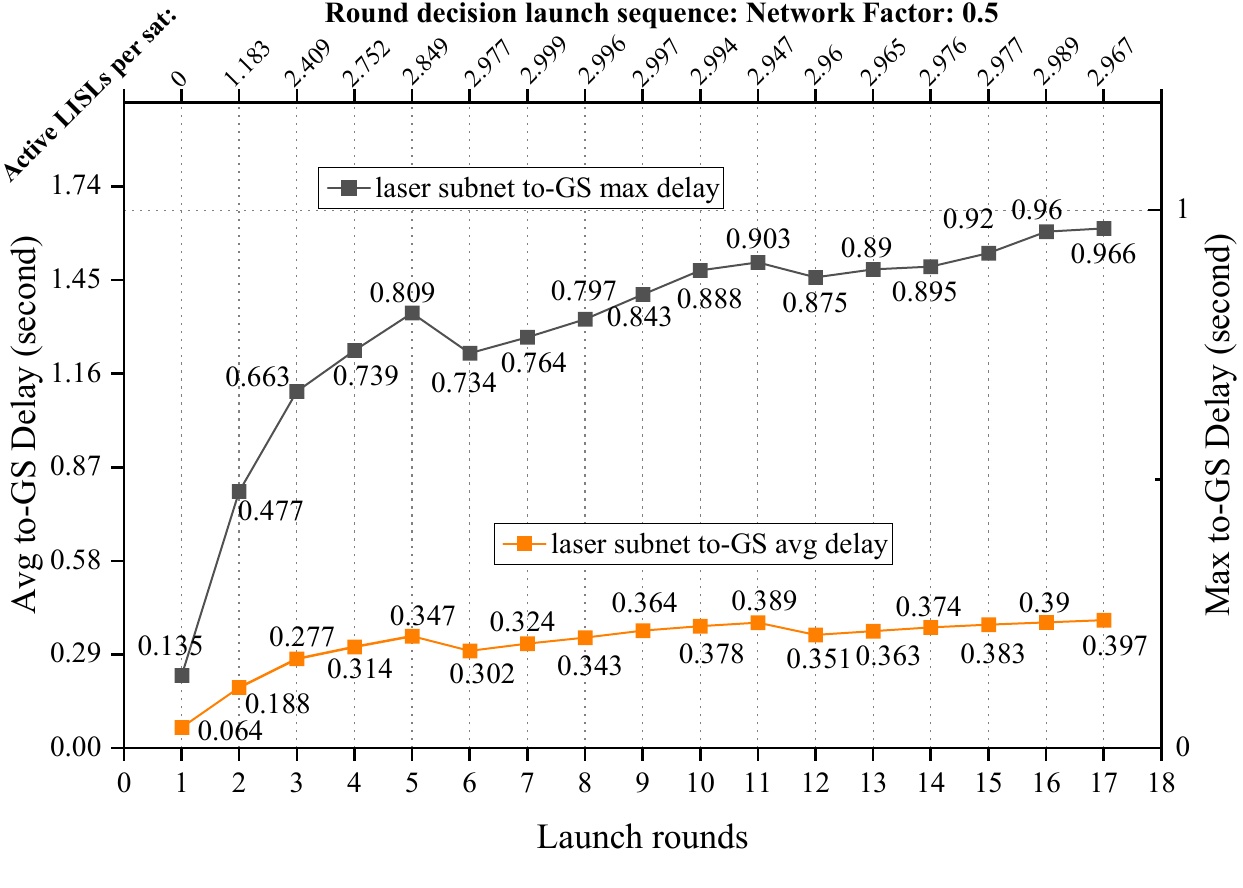}\\[-1mm]
		\textbf{(d)} $\rho=0.50$
	\end{minipage}
	\hfill
	\begin{minipage}[t]{0.3\textwidth}
		\centering
		\includegraphics[width=\linewidth]{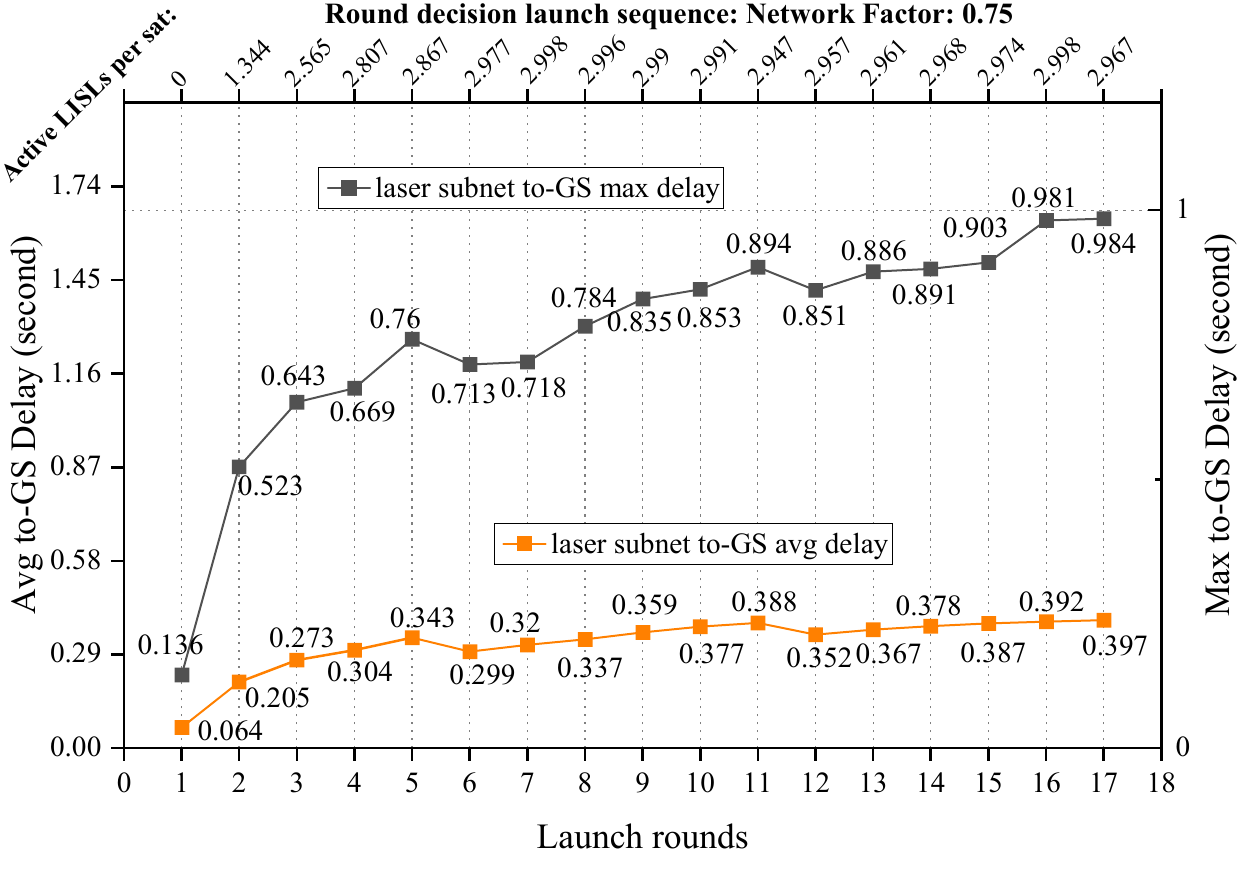}\\[-1mm]
		\textbf{(e)} $\rho=0.75$
	\end{minipage}
	\hfill
	\begin{minipage}[t]{0.3\textwidth}
		\centering
		\includegraphics[width=\linewidth]{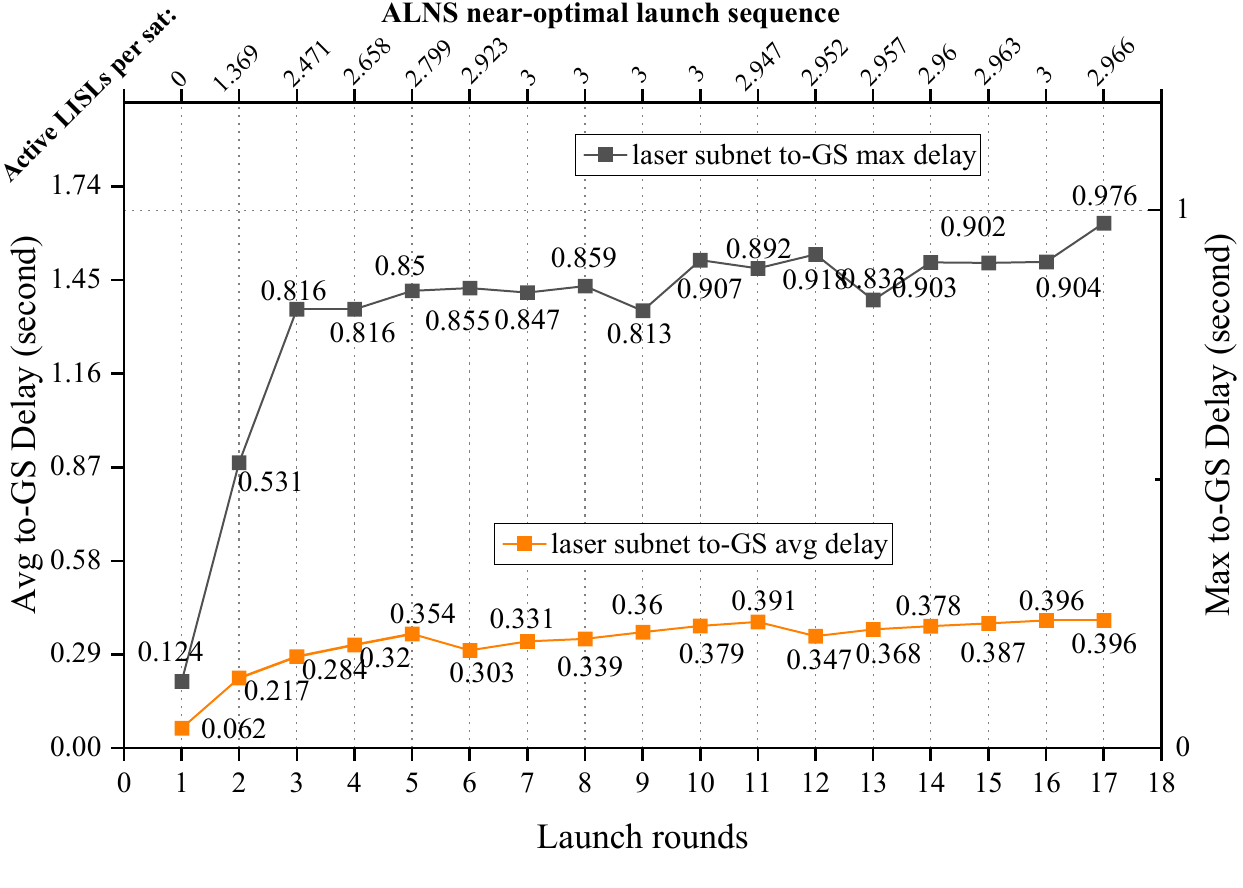}\\[-1mm]
		\textbf{(f)} ALNS
	\end{minipage}
	
	\caption{Satellite-to-ground delay and average active LISLs per satellite
		in the ground-connected laser subnetwork under the six replacement
		strategies.}
	\label{fig:laser_subnet_delay_2x3}
	
\end{figure*}

We next evaluate the satellite-to-ground delay within the ground-connected
laser subnetwork and the utilization of onboard laser terminals.
Fig.~\ref{fig:laser_subnet_delay_2x3} reports the average and maximum
satellite-to-ground delay in each launch round, together with the average
number of active LISLs per satellite. All six strategies use the same DLT
construction and three-terminal setting.

Once a ground-connected laser subnetwork is established, all strategies exhibit
a similar low-delay regime. For the mixed round-wise strategies and ALNS, the
average satellite-to-ground delay remains approximately $0.3$--$0.4$~s over
most rounds, while the maximum is generally around or below 1~s. The historical
sequence reaches a comparable level after its laser backbone becomes
ground-connected. These delays are substantially shorter than the 3-s
microwave-slot time scale, showing that the laser backbone replaces
slot-dependent store-carry-forward delivery with persistent multi-hop
forwarding dominated by propagation and transmission delay.

Laser-terminal utilization is also high. As the ground-connected laser
subnetwork grows, the average number of active LISLs per satellite approaches
$2.9$--$3.0$. With three LISL terminals available on each laser satellite,
most terminal resources are therefore used in the realized backbone during the
middle and late replacement stages.

The primary difference among the strategies is thus the time at which this
low-delay operating regime becomes available. The mixed round-wise strategies
and ALNS establish a ground-connected laser subnetwork from the first launch
round. In contrast, the historical BDS-3 sequence contains no feeder-capable
GEO/IGSO satellite during its first eight intermediate states, and the laser
subnetwork becomes ground-connected only after the first feeder-capable GEO
satellite is introduced. Hence, although the historical sequence eventually
reaches a similar subsecond delay level, the corresponding laser-network
benefit appears substantially later.

Overall, progressive replacement benefits mainly from establishing the
ground-connected laser backbone earlier and extending it to more satellites;
once formed, the backbone provides stable subsecond satellite-to-ground delay
with high utilization of the available laser terminals.
	
\subsubsection{Residual A4 Satellites Ranging Composition}
\label{subsubsec:a4_ranging_composition}

\begin{figure*}[t]
	\centering
	
	\begin{minipage}[t]{0.31\textwidth}
		\centering
		\includegraphics[width=\linewidth]{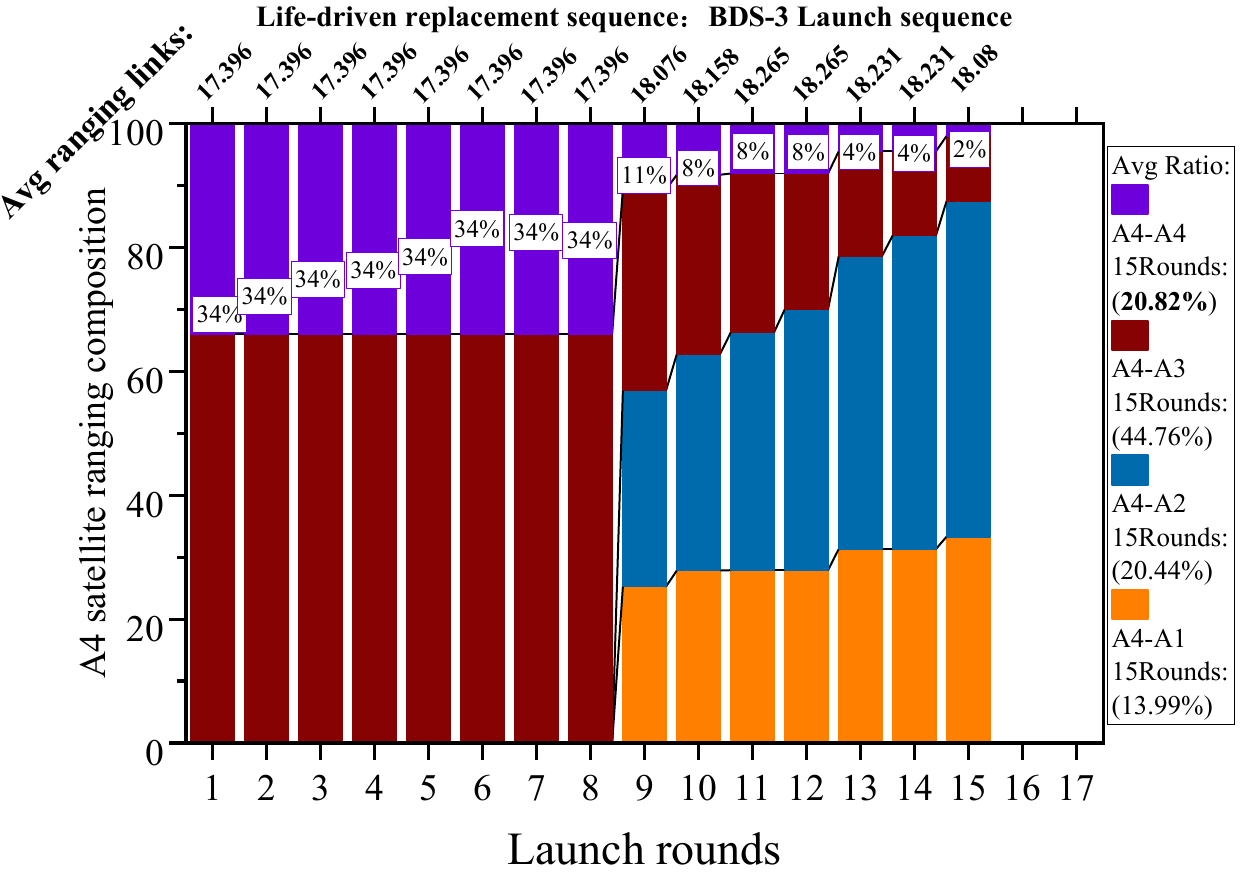}\\[-1mm]
		\textbf{(a)} Historical BDS-3
	\end{minipage}
	\hfill
	\begin{minipage}[t]{0.31\textwidth}
		\centering
		\includegraphics[width=\linewidth]{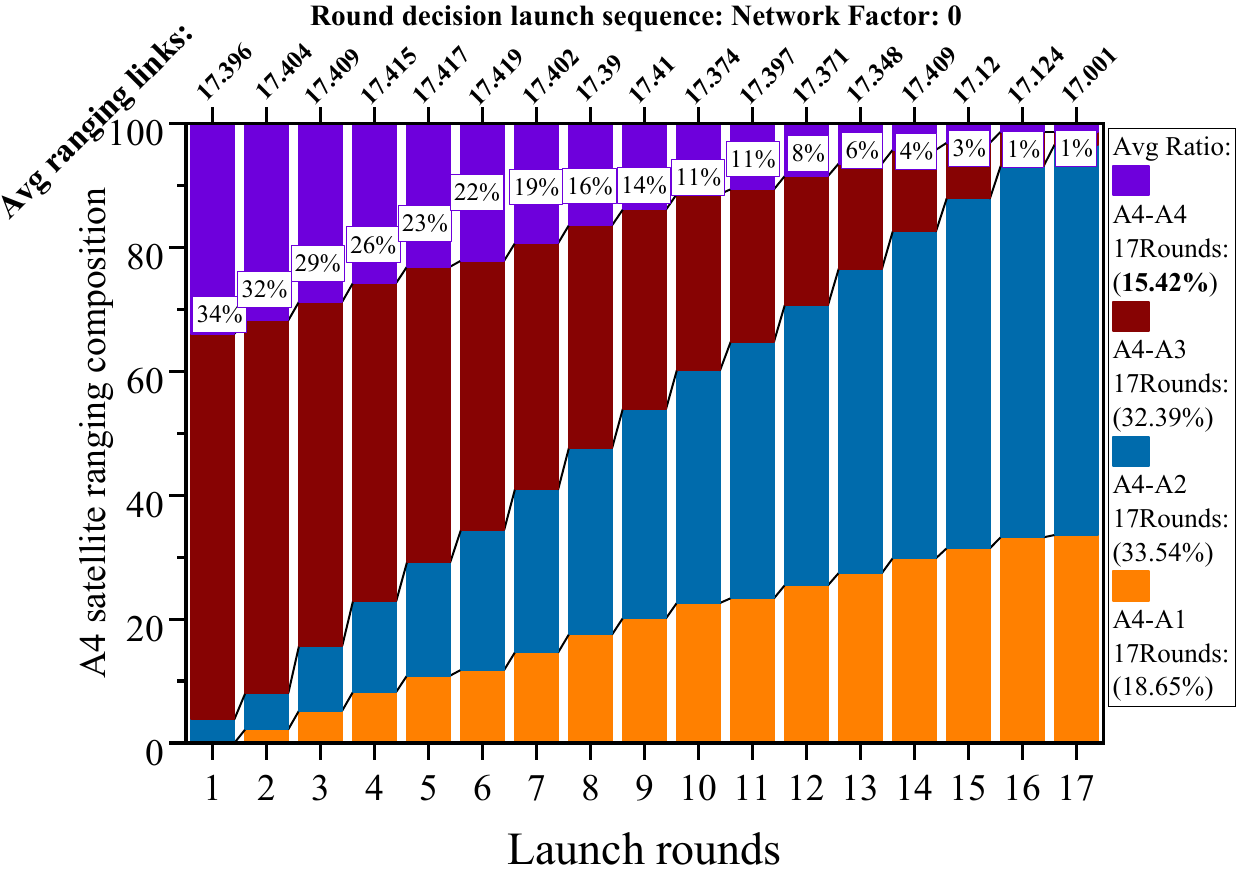}\\[-1mm]
		\textbf{(b)} $\rho=0$
	\end{minipage}
	\hfill
	\begin{minipage}[t]{0.31\textwidth}
		\centering
		\includegraphics[width=\linewidth]{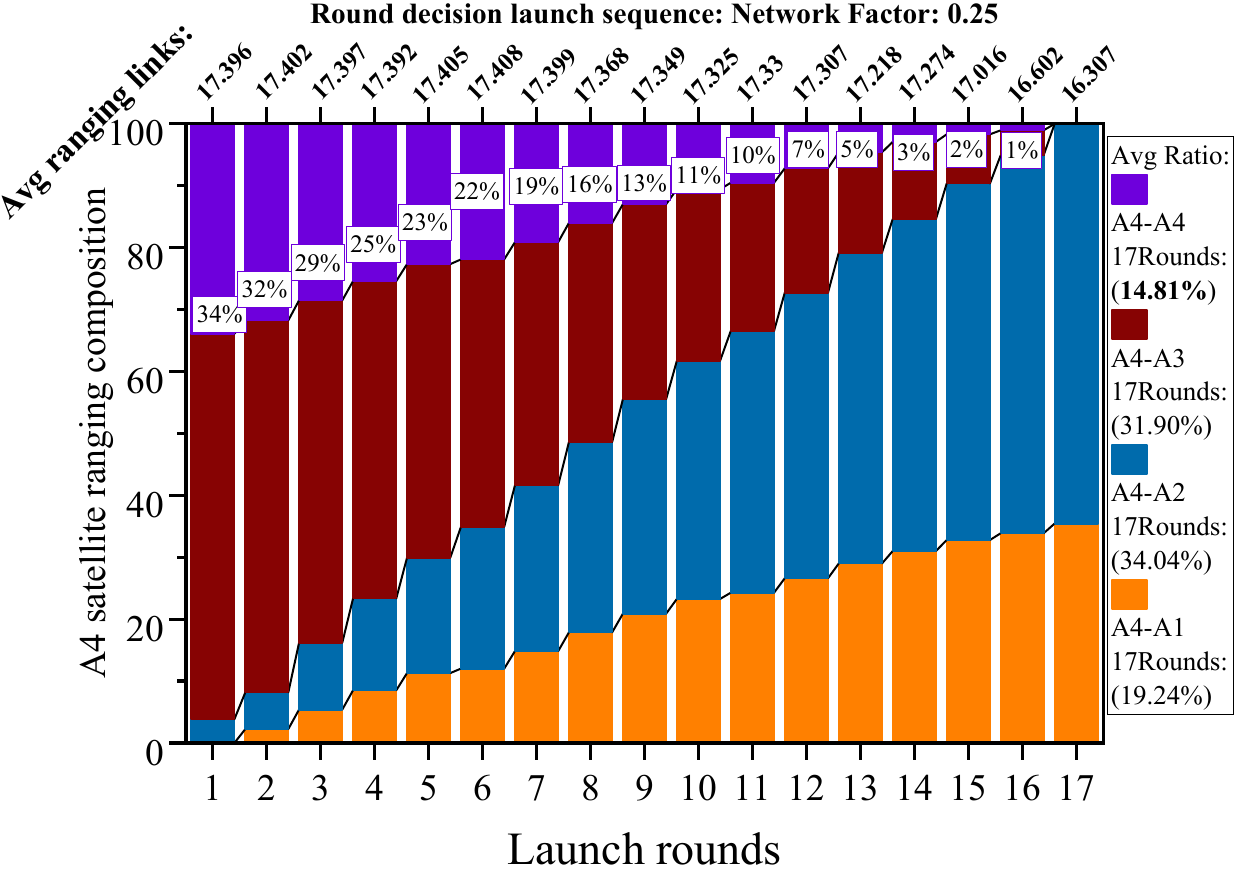}\\[-1mm]
		\textbf{(c)} $\rho=0.25$
	\end{minipage}
	
	\vspace{2mm}
	
	\begin{minipage}[t]{0.31\textwidth}
		\centering
		\includegraphics[width=\linewidth]{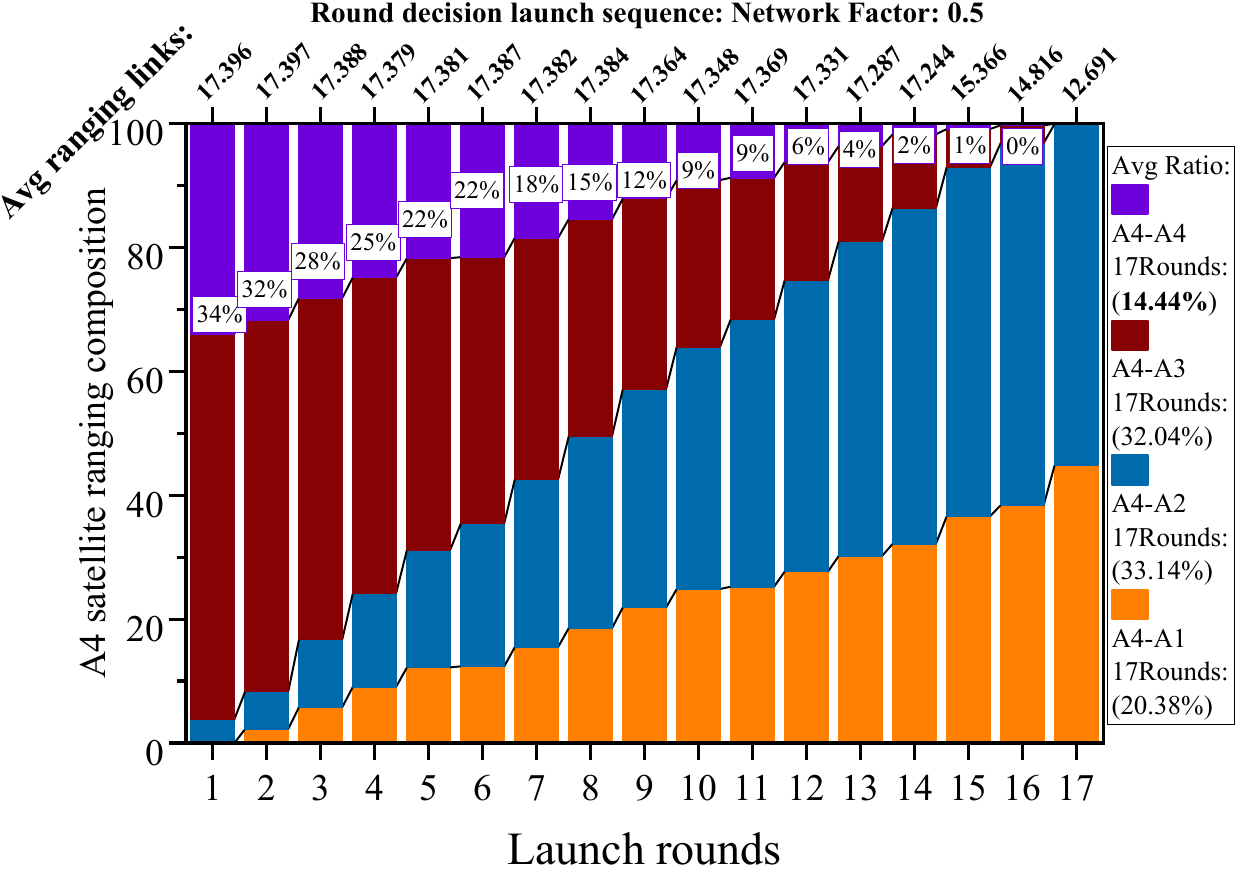}\\[-1mm]
		\textbf{(d)} $\rho=0.50$
	\end{minipage}
	\hfill
	\begin{minipage}[t]{0.31\textwidth}
		\centering
		\includegraphics[width=\linewidth]{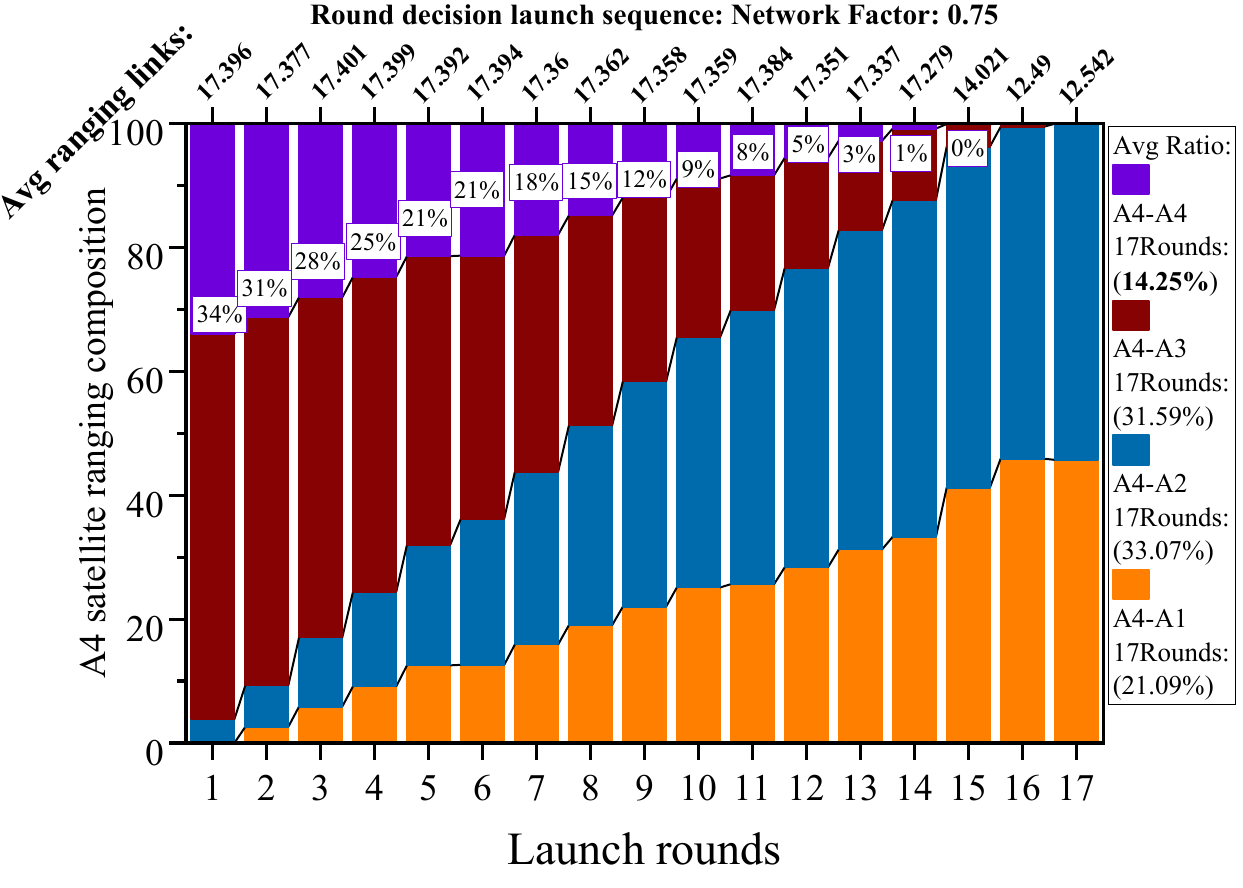}\\[-1mm]
		\textbf{(e)} $\rho=0.75$
	\end{minipage}
	\hfill
	\begin{minipage}[t]{0.31\textwidth}
		\centering
		\includegraphics[width=\linewidth]{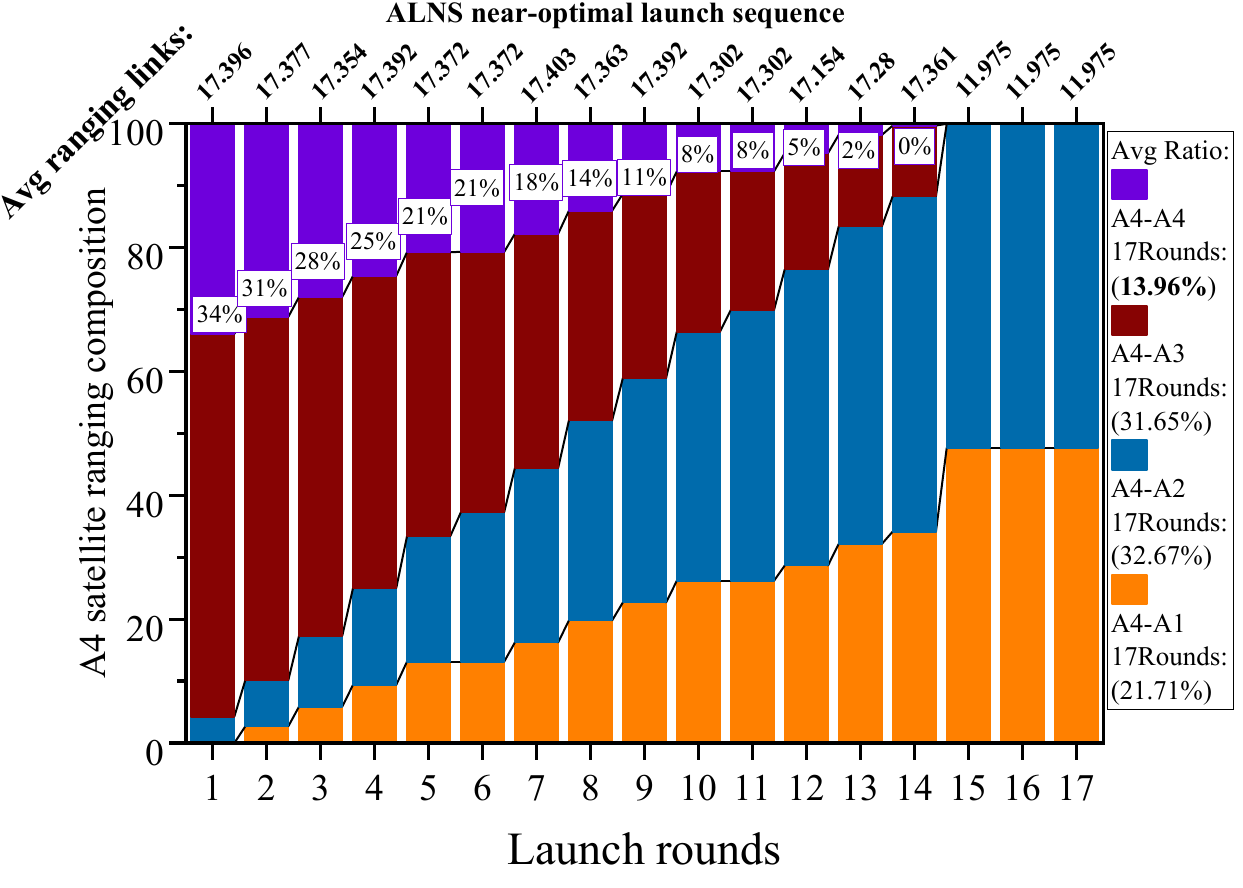}\\[-1mm]
		\textbf{(f)} ALNS
	\end{minipage}
	
	\caption{Ranging-link composition of A4 satellites under the six
		replacement strategies.}
	\label{fig:a4_ranging_2x3}
\end{figure*}

After the growth of the ground-connected laser backbone, the remaining
performance bottleneck is concentrated on A4 satellites. These satellites are
outside the laser subnetwork and are also invisible to the ground segment, so
their measurement and communication opportunities depend on microwave contacts
with other satellites. Fig.~\ref{fig:a4_ranging_2x3} reports the ranging-link
composition of A4 satellites under the six replacement strategies. The stacked
bars distinguish A4--A4, A4--A3, A4--A2, and A4--A1 ranging relations, and the
upper axis gives the total number of A4 ranging links in each round.

A clear structural trend appears as network participation increases. The
fraction of A4--A4 ranging links decreases progressively from the
engineering-dominated cases to the higher-$\rho$ strategies and further to
ALNS, while the fractions of A4--A1, A4--A2, and A4--A3 links increase. This
shift means that the residual A4 satellites establish fewer measurements with
other non-anchor satellites and more measurements with the anchor set. From the
communication perspective, such a contact pattern gives A4 satellites more
direct opportunities to reach satellites that can already access the ground.
From the timing perspective, it shortens the path from A4 satellites to the
ground-synchronized part of the constellation. In particular, the increase of
A4--A1 links directly reflects the expansion of the ground-connected laser
backbone, because A1 satellites are precisely the ground-invisible satellites
that have been converted into anchor satellites through laser connectivity.

The difference between the historical BDS-3 sequence and the network-assisted
strategies is again pronounced in the early and middle replacement stages.
Under the historical sequence, the delayed introduction of feeder-capable
GEO/IGSO satellites postpones the formation of the ground-connected laser
backbone. As a result, the A4--A4 component remains relatively large for a long
period, and the shift toward anchor-related ranging relations occurs only after
the laser backbone becomes ground-connected. By contrast, the mixed round-wise
strategies and ALNS drive the ranging composition toward A4--A1/A2/A3 contacts
much earlier, which is consistent with the earlier backbone formation observed
in Fig.~\ref{fig:laser_subnet_size_2x3}.

The total number of A4 ranging links shows an additional feature in the late
replacement rounds. In the higher-$\rho$ strategies and ALNS, the total A4
ranging-link count can become slightly smaller than that of the lower-$\rho$
strategies. This behavior comes from the identity of the remaining A4
satellites rather than from a degradation of the network structure. Network
participation tends to postpone the launch of IGSO1 to the final rounds,
because a single-satellite IGSO action contributes less immediately to
laser-backbone expansion and, as shown by the ground-visibility statistics in
Fig.~\ref{fig:gs_visibility_fsa_count}, IGSO1 is also not the most persistent
choice for forming high-rate ground access. When the A4 population has already
become very small, IGSO1 can account for a large fraction of the residual A4
set. Since its microwave visibility degree is only about 10--12 satellites, the
resulting total number of A4 ranging links is correspondingly lower. This
effect is therefore a normal consequence of the residual-satellite composition
in the late rounds.

Overall, stronger network participation gives the remaining A4 satellites a
more favorable ranging environment: fewer A4--A4 contacts, more direct ranging
relations with anchor satellites, and a faster transition of the residual
microwave layer toward anchor-oriented interaction.

\subsubsection{Residual A4 Satellites Waiting Delay}
\label{subsubsec:a4_waiting_delay}

\begin{figure*}[t]
	\centering
	
	\begin{minipage}[t]{0.3\textwidth}
		\centering
		\includegraphics[width=\linewidth]{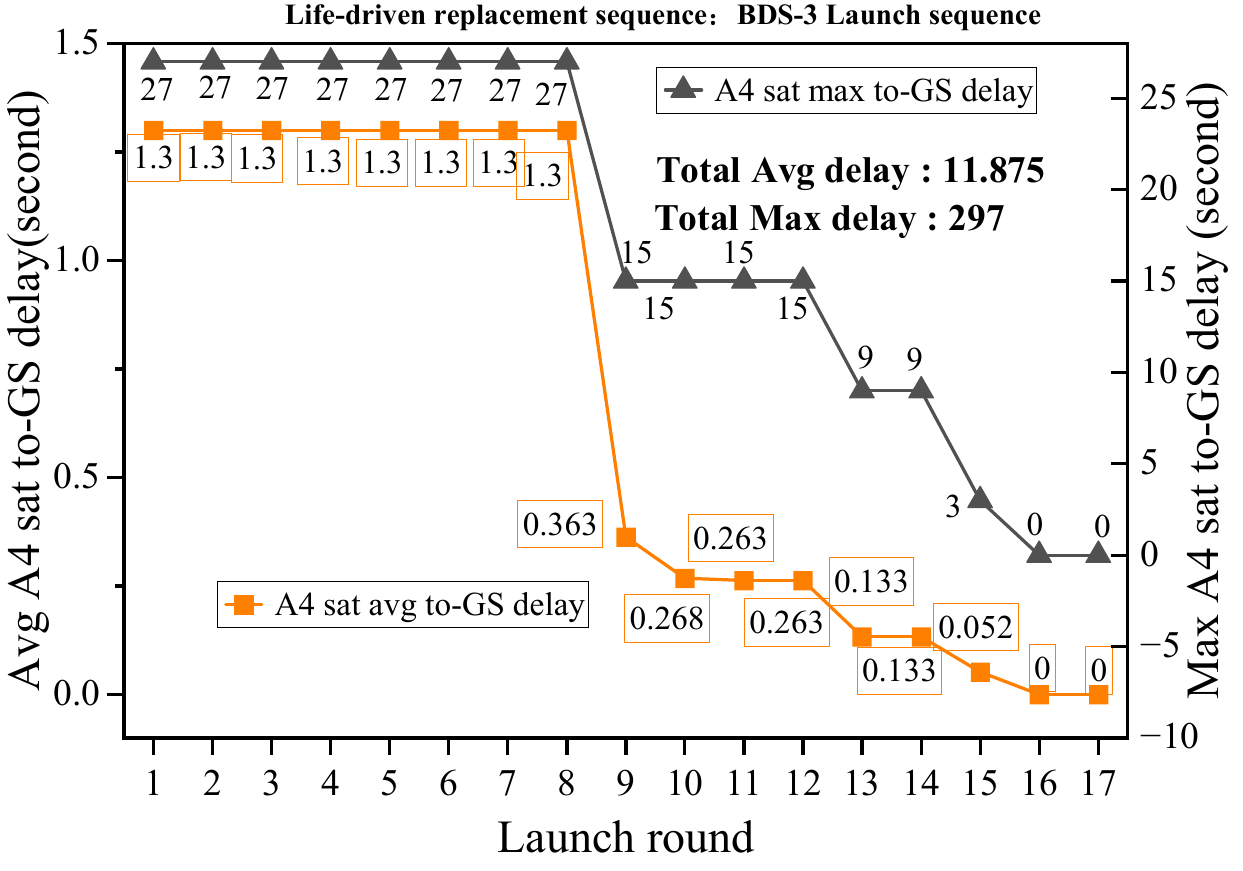}\\[-1mm]
		\textbf{(a)} Historical BDS-3
	\end{minipage}
	\hfill
	\begin{minipage}[t]{0.3\textwidth}
		\centering
		\includegraphics[width=\linewidth]{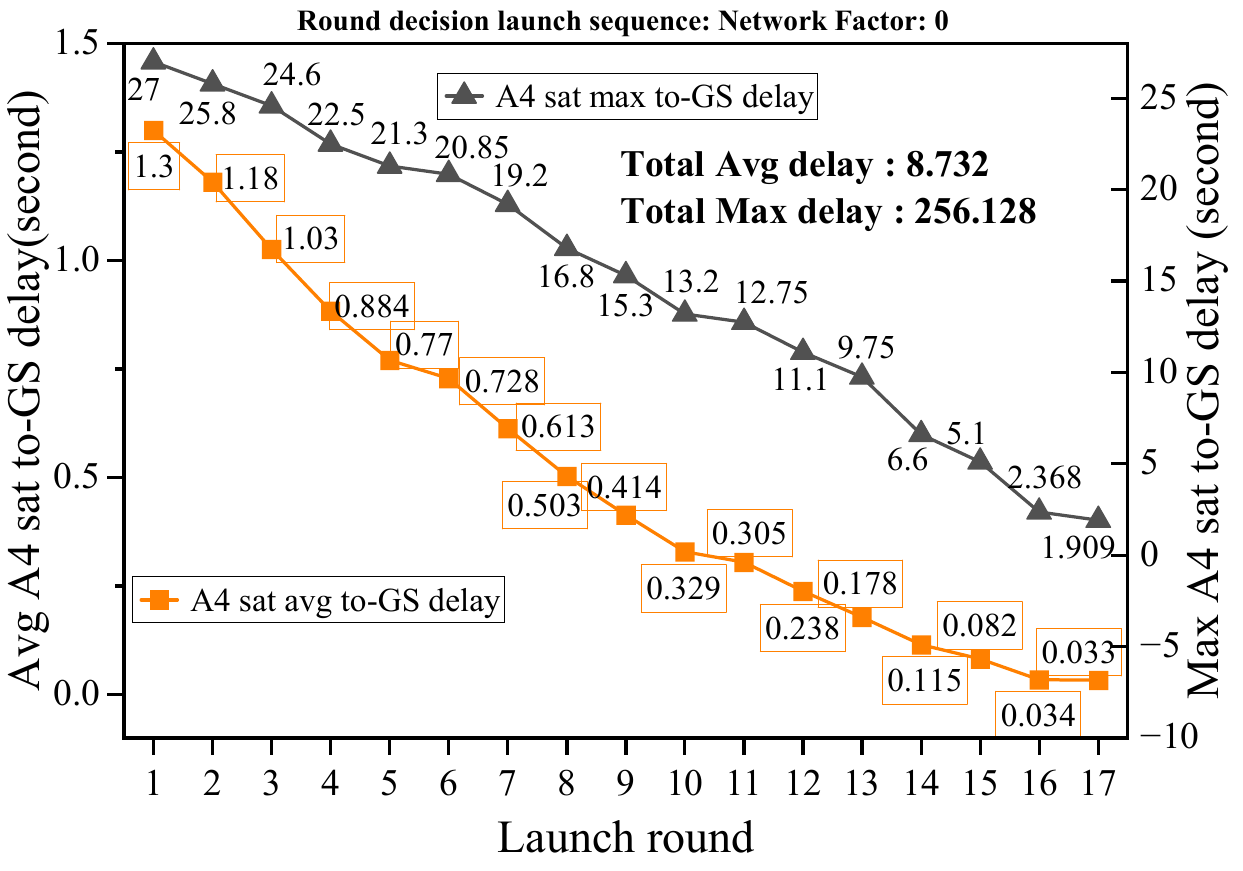}\\[-1mm]
		\textbf{(b)} $\rho=0$
	\end{minipage}
	\hfill
	\begin{minipage}[t]{0.3\textwidth}
		\centering
		\includegraphics[width=\linewidth]{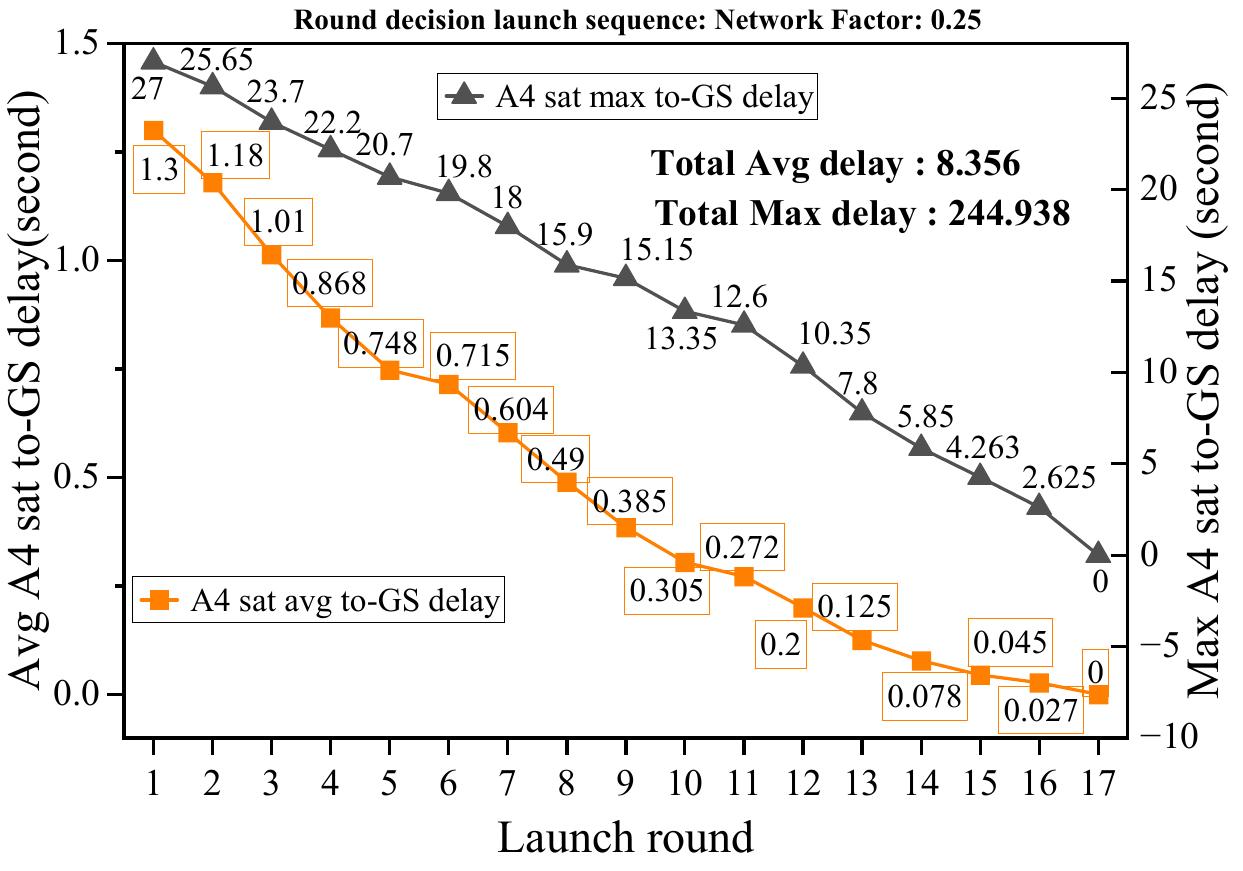}\\[-1mm]
		\textbf{(c)} $\rho=0.25$
	\end{minipage}
	
	\vspace{2mm}
	
	\begin{minipage}[t]{0.3\textwidth}
		\centering
		\includegraphics[width=\linewidth]{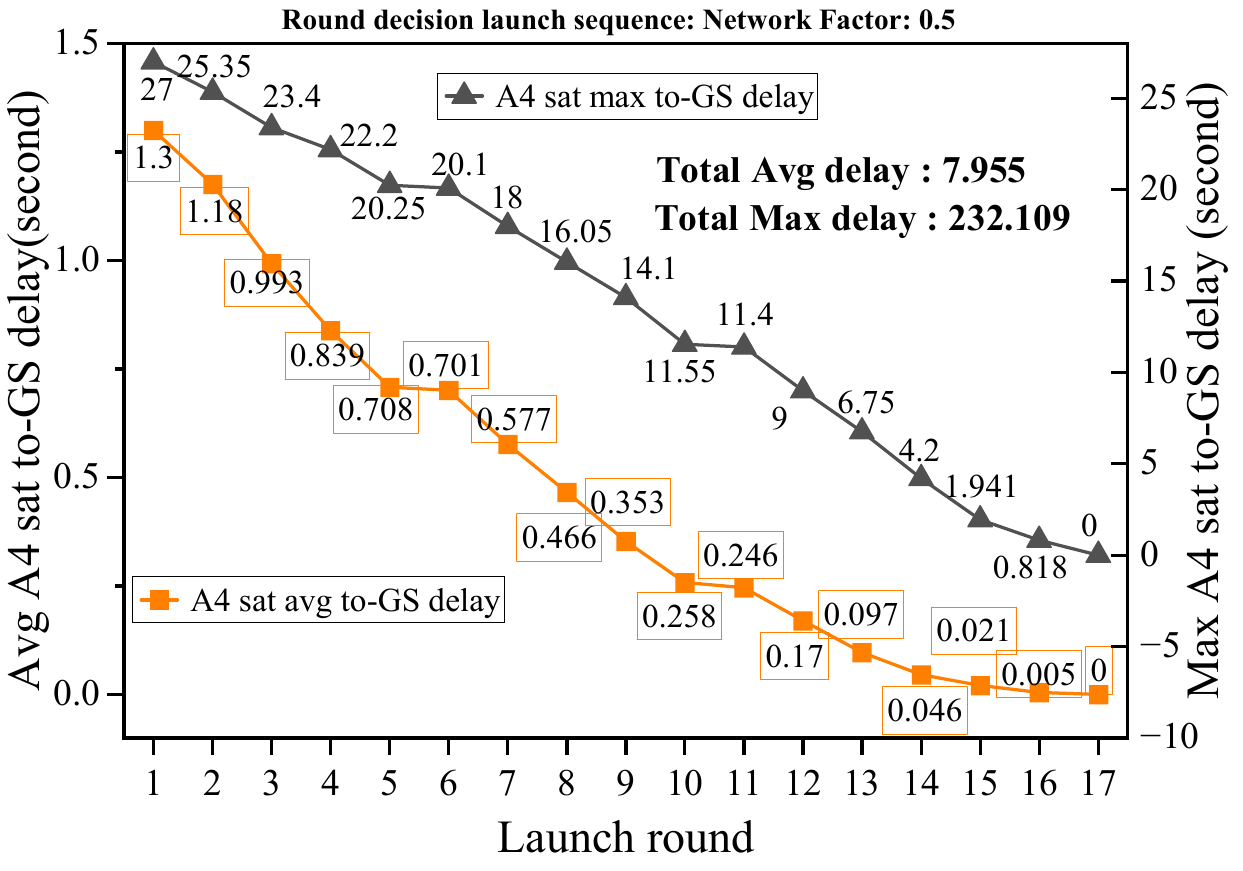}\\[-1mm]
		\textbf{(d)} $\rho=0.50$
	\end{minipage}
	\hfill
	\begin{minipage}[t]{0.3\textwidth}
		\centering
		\includegraphics[width=\linewidth]{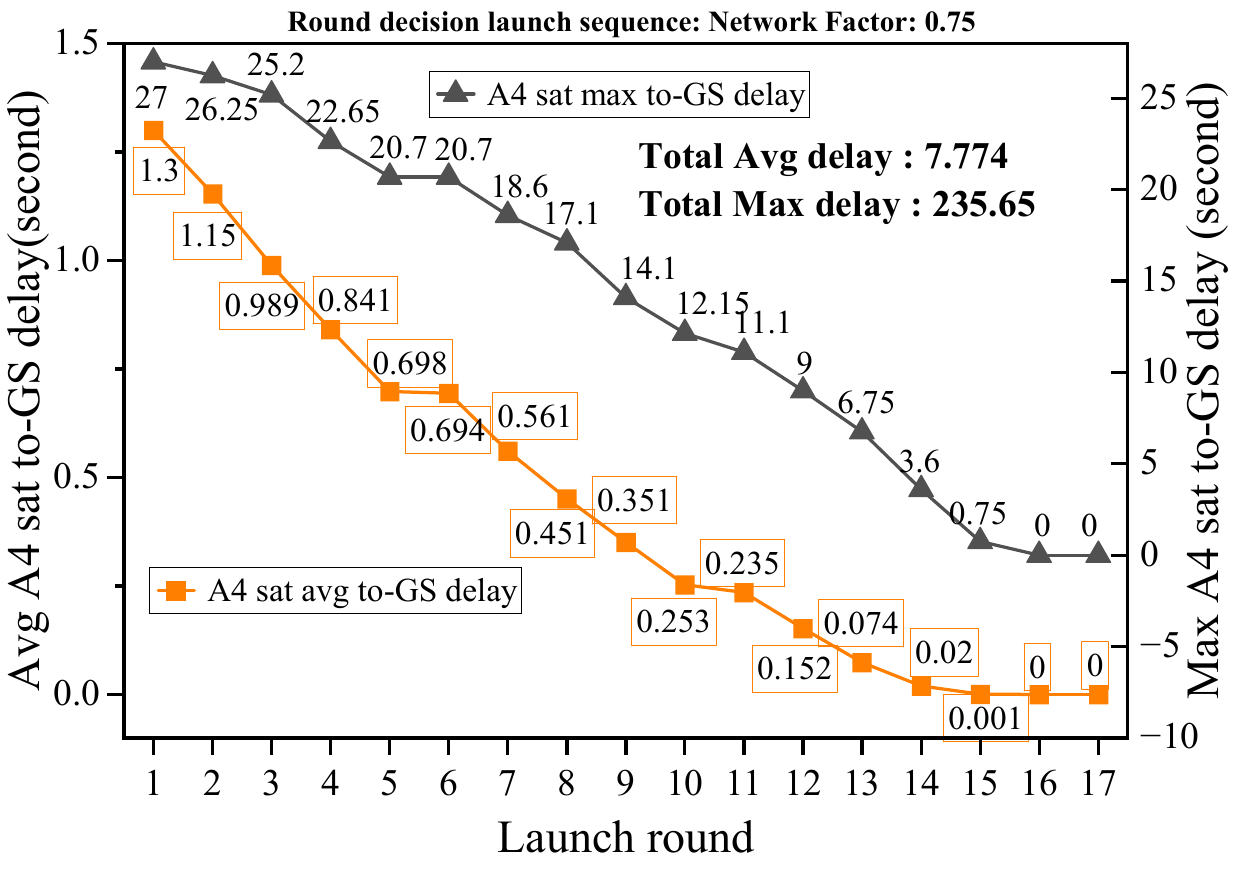}\\[-1mm]
		\textbf{(e)} $\rho=0.75$
	\end{minipage}
	\hfill
	\begin{minipage}[t]{0.3\textwidth}
		\centering
		\includegraphics[width=\linewidth]{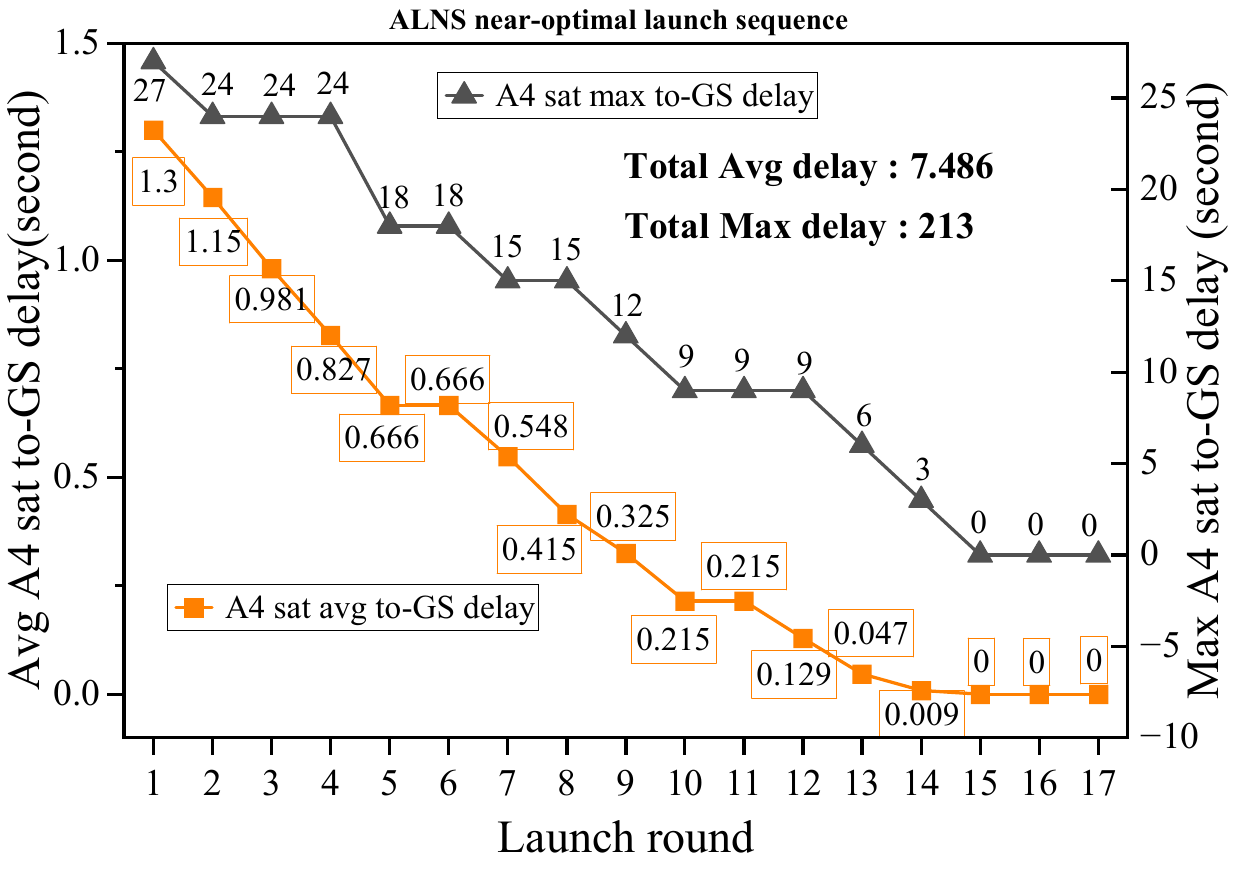}\\[-1mm]
		\textbf{(f)} ALNS
	\end{minipage}
	
	\caption{Average and maximum waiting delay for A4 satellites to access the
		anchor set under the six replacement strategies.}
	\label{fig:a4_delay_2x3}
\end{figure*}

The communication performance of residual A4 satellites is further evaluated by
their waiting delay before obtaining a microwave contact opportunity to the
anchor set. Fig.~\ref{fig:a4_delay_2x3} reports the average and maximum A4
waiting delay under the six replacement strategies. The reported delay is the
slot-level DTN waiting delay in the microwave scheduling layer. It quantifies
the time that an A4 satellite waits before accessing an anchor satellite and
does not include propagation delay, whose contribution is secondary on this
time scale.

The average A4 waiting delay decreases as the network participation ratio
increases. Under the higher-$\rho$ strategies and ALNS, the average delay drops
more rapidly in the middle replacement rounds and remains lower thereafter.
This result follows directly from the structural evolution reported in Fig.~\ref{fig:laser_subnet_size_2x3}. A larger ground-connected laser backbone creates more A1
anchor satellites, the total anchor set expands, and the number of residual A4
satellites decreases. The remaining A4 satellites therefore obtain microwave
contacts to the anchor set more frequently, which shortens their waiting time
before data return and timing access.

The maximum waiting delay exhibits a more fluctuating behavior. This metric is
determined by the most disadvantaged A4 satellite in each round and is
therefore sensitive to the instantaneous contact opportunities of a small
number of residual nodes. As the A4 population becomes small, one remaining
difficult satellite can dominate the round-wise maximum. The presence of such
fluctuations does not change the overall trend: strategies with stronger
network participation still drive the maximum delay downward earlier and bring
it to zero in fewer remaining rounds as the A4 set disappears.

The historical BDS-3 sequence again improves later than the network-assisted
strategies. Because the ground-connected laser backbone forms late, the anchor
set expands slowly in the early replacement rounds, and the A4 waiting delay
remains high for a longer interval. Once feeder-capable GEO/IGSO satellites are
introduced and the laser backbone begins to support more anchors, the delay
starts to decrease. The mixed round-wise strategies and ALNS reach this delay
reduction much earlier, showing that earlier backbone formation improves not
only the size of the anchor set but also the communication experience of the
residual microwave satellites.

Taken together, Fig.~\ref{fig:a4_ranging_2x3} and
Fig.~\ref{fig:a4_delay_2x3} show a consistent residual-layer trend. As network
participation deepens, the remaining A4 satellites range less frequently with
other A4 satellites and more frequently with anchors, while their DTN waiting
delay to the anchor set decreases accordingly. This improvement follows from
both levels of the networking potential objective. The primary component
expands the ground-connected laser subnetwork, converting more
ground-invisible satellites into A1 anchors and reducing the residual A4
population. At the attained backbone scale, the secondary component further
maximizes the microwave-visibility support from the laser subnetwork to the
residual microwave segment. Consequently, the remaining A4 satellites obtain
more direct contact opportunities with the anchor set, which shifts the ranging
composition away from A4--A4 links and shortens the waiting time before
accessing the laser-enabled ground-return domain. The residual-layer ranging
and delay improvements therefore directly reflect the hierarchical
networking potential model used in replacement planning.

	\subsection{Discussion}
	\label{subsec:evaluation_discussion}
	
	The results establish a consistent relation between network participation in
	launch decisions and the quality of the intermediate hybrid constellation
	states. As $\rho$ increases, the replacement process forms a larger
	ground-connected laser subnetwork, leaves fewer satellites in the residual A4
	segment, and achieves closer matching between laser-backbone growth and
	high-rate ground access. The residual microwave segment also shifts toward
	anchor-oriented ranging and experiences lower average waiting delay. Once a
	ground-connected laser subnetwork is formed, its satellite-to-ground delay
	remains at a similar subsecond level across the considered strategies, with
	high utilization of the available laser terminals. The main benefit of
	network-aware sequence planning therefore lies in activating this high-quality
	laser domain earlier and extending it to a larger portion of the constellation
	throughout the replacement process.
	
	In an actual replacement campaign, launch decisions are jointly shaped by
	remaining lifetime, on-orbit condition, satellite production, launch-task
	arrangement, and network evolution. Engineering conditions first determine the
	admissible replacement actions, and networking potential then provides an
	additional criterion for distinguishing among these feasible choices. The
	network-participation ratio $\rho$ captures this role through the fraction of
	launch rounds in which networking information enters the decision process.
	The progressive improvement from $\rho=0$ to $\rho=0.75$ shows that network
	planning can contribute useful gains even when it participates in only part of
	the campaign. This makes round-wise planning suitable for replacement
	processes in which engineering conditions are updated from one launch
	opportunity to the next.
	
	When networking information participates in the complete launch process, the
	decision scope naturally expands from individual launch opportunities to the
	entire sequence of intermediate states. The ALNS result represents this
	full-horizon endpoint and provides the strongest overall hybrid-network
	evolution among the considered strategies. Together, the round-wise and
	full-sequence results show two complementary uses of the proposed framework:
	networking potential can support selected engineering decisions during an
	ongoing replacement campaign, and it can also guide full-sequence planning
	when complete campaign information is available. In both cases, the central
	planning objective remains the same: shaping the launch order so that the
	constellation obtains the benefits of the laser-enabled network as early and
	as extensively as possible during progressive replacement.

	\section{Conclusion}
	\label{sec:conclusion}
	
	This paper investigated progressive replacement-sequence planning for the
	BeiDou navigation constellation from a networking perspective. Network
	performance is treated as an auxiliary criterion within the actual engineering
	replacement process. A common networking potential measure is established for
	intermediate laser--microwave hybrid states, and two complementary planning
	modes are developed: priority-aware exact round-wise planning for
	network-assisted decisions at individual launch opportunities, and
	visibility-guided ALNS for full-sequence planning from a full-horizon
	perspective. An SCF-ILP model provides the common state evaluator for both
	modes, while priority-aware search and deployed-state reuse improve
	sequence evaluation efficiency.
	
	Evaluation with the actual BDS-3 constellation shows that stronger network
	participation consistently improves the intermediate hybrid states. It enables
	earlier and larger ground-connected laser subnetworks, reduces the residual A4
	population, and improves the coordination between laser-backbone growth and
	high-rate ground access. Once formed, the laser backbone provides subsecond
	satellite-to-ground delay with high laser-terminal utilization. The residual
	microwave layer also benefits from fewer A4--A4 ranging relations, more direct
	contacts with anchor satellites, and lower average DTN waiting delay.
	
	These results show that networking information can provide measurable value
	even when it participates in only part of the replacement process. Round-wise
	planning therefore supports network-aware engineering decisions as feasible
	launch choices evolve, while full-sequence ALNS provides a near-optimal
	full-horizon reference when networking considerations extend across the entire
	campaign. The proposed framework thus offers a unified
	basis for incorporating network performance into future BeiDou replacement
	planning and for shaping the launch order so that laser-enabled networking
	capabilities become available earlier and serve a larger portion of the
	constellation throughout progressive replacement.
	
	\bibliographystyle{IEEEtran}
	\bibliography{references}
		
	\section{Funding Declaration}
	\label{sec:funding}
	This work is supported in part by the National Natural Science Foundation of China under Grant 62131012 and the Unveiling and Leading Project of Nanjing University Integrated Research and Development Platform of Ministry of Education, in part by the Nanjing University and Skolkovo Institute of Science and Technology International Joint Research Center for STEM Collaborative Research Grant 2025 (LunarCubeNet).

 	\appendix[Historical BDS-3 and ALNS Replacement Sequences]
 \label{app}
 \begin{table}[H]
 	\centering
 	\caption{Historical BDS-3 and ALNS Replacement Sequences}
 	\label{tab:replacement_sequences_appendix}
 	\footnotesize
 	\setlength{\tabcolsep}{4pt}
 	\begin{tabular}{c c c}
 		\toprule
 		Round & Historical BDS-3 & ALNS \\
 		\midrule
 		1  & M1, M2   & GEO2 \\
 		2  & M3, M4   & M8, M22 \\
 		3  & M5, M6   & M6, M18 \\
 		4  & M7, M8   & M7, M16 \\
 		5  & M9, M10  & M12, M24 \\
 		6  & M11, M12 & IGSO2 \\
 		7  & M13, M14 & M1, M14 \\
 		8  & M15, M16 & M4, M19 \\
 		9  & GEO1     & M11, M17 \\
 		10 & M17, M18 & M5, M23 \\
 		11 & IGSO1    & IGSO3 \\
 		12 & IGSO2    & M9, M21 \\
 		13 & M23, M24 & M10, M15 \\
 		14 & IGSO3    & M13, M2 \\
 		15 & M21, M22 & M20, M3 \\
 		16 & M19, M20 & GEO3 \\
 		17 & GEO2     & GEO1 \\
 		18 & GEO3     & IGSO1 \\
 		\bottomrule
 	\end{tabular}
 \end{table}
 For reproducibility, Table~\ref{tab:replacement_sequences_appendix}
 lists the complete historical BDS-3 replacement sequence and the
 best-found full-sequence ALNS result used in the evaluation. Both sequences
 follow the launch-grouping rule adopted in this paper: same-plane MEO
 satellites are launched in pairs, whereas GEO and IGSO satellites are launched
 individually. The 18th action completes constellation replacement; therefore,
 only the first 17 post-launch states contribute to the inter-launch
 sequence-level objective.


\end{document}